\documentclass{aa} 

\usepackage{graphicx}
\usepackage{adjustbox}
\usepackage{natbib}
\usepackage[T1]{fontenc}
\usepackage{ae,aecompl}
\usepackage{rotating}
\usepackage{appendix}
\usepackage{longtable}
\usepackage{xurl}
\usepackage[bookmarksnumbered]{hyperref}
\usepackage{txfonts}
\usepackage{hyperref}

\begin{document} 

\titlerunning{New ULXs in NGC 4631 and NGC 1097}
\authorrunning{S. Allak et al.}

\title{Identification of three new ultraluminous X-ray sources in NGC 4631 and NGC 1097: 
Evidence for stellar-mass black holes}

\author{Sinan Allak\inst{1}\thanks{Corresponding author: 
\email{sinan.allak@uni-tuebingen.de}}, 
Aysun Akyuz\inst{2,3}, 
Yasemin Aladag\inst{2}, 
Lorenzo Ducci\inst{1}, 
and Andrea Santangelo\inst{1}
}

\institute{Institut für Astronomie und Astrophysik, Sand 1, 72076 
Tübingen, Germany\\
\email{sinan.allak@uni-tuebingen.de}
\and
Department of Physics, University of Çukurova, 01330, Adana, Türkiye
\and
Space Science and Solar Energy Research and Application Center (UZAYMER), 
University of Çukurova, 01330, Adana, T\"urkiye
}

 \date{Received September 15, 2025; accepted .. .., ..}

 \abstract
 
{Recent observations of galaxies continue to reveal new ultraluminous X-ray sources (ULXs), thereby increasing their known population. This growing sample provides improved statistics that are essential for advancing our understanding of ULX properties and their nature.
}
 {Our objective was to study the source populations using published \textit{Chandra}, \textit{XMM-Newton}, and \textit{Swift}/XRT observations to identify new ULXs. In particular, we focused on investigating their X-ray and optical properties to constrain the nature of the compact objects and donor stars.}
 {We analyzed archival X-ray observations spanning 2000 to 2025 for NGC 4631 and NGC 1097 to study their ULX populations. We performed spectral fitting for sources with sufficient counts using absorbed power-law and diskbb models to determine their physical properties. We conducted variability analyses, including hardness-intensity diagrams and light curves, to assess both short- and long-term variability. Additionally, we used optical color-magnitude diagrams and near-infrared (NIR) spectral energy distributions (SEDs) to identify and characterize possible donor stars.}
 {In NGC 4631, we identify two new transient ULXs, X-6 and X-7, whose X-ray count rates vary by more than an order of magnitude. The $L_{X} \propto T^{4}$ relation obtained from the diskbb fits provides strong evidence that NGC 4631 X-6 is powered by a stellar-mass black hole (BH) accreting via a standard disk. The optical sources within the X-6 and X-7 X-ray error circles are candidate optical counterparts, suggesting that these systems are candidate high-mass X-ray binaries. In NGC 1097, we report the discovery of a new transient ULX, designated ULX-3, which exhibits variations in X-ray luminosity (L$_{X}$) by a factor of about 30. For ULX-3, we detect spectral state transitions, which may indicate a compact object consistent with either a stellar-mass BH or a neutron star. Moreover, we identify a unique optical and NIR counterpart: while the optical emission is variable, the NIR emission remains stable. The NIR counterpart of ULX-3 shows an SED consistent with a blackbody temperature of $\sim3300$ K, which lies within the expected range for red supergiants with a stellar radius of $\sim200 R_{\odot}$.}


 \maketitle

\section{Introduction}

Ultraluminous X-ray sources (ULXs) are the most extreme subclass of X-ray binaries (XRBs), appearing as off-nuclear point sources exhibiting $L_{\rm X}$ > 2$\times$$10^{39}$ erg s$^{-1}$ \citep{2017ARA&A..55..303K,2021AstBu..76....6F,2023NewAR..9601672K,2023arXiv230200006P}. To account for their ultraluminosities, recent studies increasingly support the view that the majority of ULXs are powered by stellar-mass compact objects undergoing super-Eddington accretion (\cite{2023NewAR..9601672K} and references therein), rather than by sub-Eddington accretion onto intermediate-mass black holes\citep{Heinzeller2006, 2019BAAS...51c.352B, 2022cosp...44.2201G}. Within this framework, the accreting compact object is generally inferred to be either a black hole (BH) or a neutron star (NS).

Coherent pulsations have been detected in several ULXs, confirming them as pulsating NS ULXs (PULXs), primarily through observations by \textit{XMM-Newton} and \textit{NuSTAR} \citep{2014Natur.514..202B,2014Natur.514..198M,2017Sci...355..817I,2018MNRAS.476L..45C}.  Moreover, the observation of cyclotron resonance scattering features in the X-ray spectrum of a ULX implies the presence of an NS candidate \citep{2017MNRAS.467.1202M, 2020MNRAS.491.5702P}. 
In addition, several ULXs, including some identified as PULXs, show transient behavior, whereas most remain persistently bright in the X-ray band over long timescales. Repeated observations of nearby galaxies with \textit{Chandra} and \textit{XMM-Newton} have contributed to a growing number of ULXs, particularly those exhibiting transient behavior \citep{2019MNRAS.483.3566V,2020ApJ...891..153E,2021MNRAS.501.1002W,2022MNRAS.515.4669R,2023MNRAS.525.3330R,2023ApJ...951...51B,2023MNRAS.526.5765A}.

Several ULXs also show clear evidence of recurrent spectral state transitions, evolving from hard to soft regimes over time, as revealed by high-cadence and long-term studies. In particular, Holmberg II X1 and NGC 5204 X1 show a cyclical evolution through the states of hard ultra-luminous  (HUL), soft ultra-luminous (SUL), and supersoft ultra-luminous (SSUL), likely driven by variations in accretion rate and changes in funnel geometry caused by radiation‑driven winds \citep{2021A&A...654A..10G,2021A&A...651A..54M}. A recent study by \cite{2025MNRAS.539.2064M} analyzed the long-term evolution of several ULXs with candidate BH accretors using \textit{XMM-Newton}, revealing diverse spectrotemporal behaviors and correlations between disk luminosity (L$_\mathrm{disc}$) and inner-disk temperature (T$_\mathrm{col}$). The authors argue that these correlations are indicative of possible spectral states or transitions between standard thin-disk and slim-disk configurations at different luminosities, supporting a two-component supercritical accretion framework regulated by the disk, corona, and radiation-driven winds.
Numerous studies have sought to identify the optical counterparts of ULXs and clarify the origin of their emission \citep{2011ApJ...737...81T,2012MNRAS.420.3599S,2013ApJS..206...14G,2018ApJ...854..176V,2022MNRAS.515.3632A,2023ApJ...946...72G,2025A&A...694A.301A}. However, determining the precise contributions remains challenging, as the emission may originate from the donor star’s photosphere, the accretion disk, or a combination of both \citep{2011ApJ...737...81T,2015NatPh..11..551F,2022MNRAS.515.3632A}. In rare cases, spectral signatures such as stellar absorption lines allow clear donor identification, as seen in NGC 7793 P13, where a B9Ia supergiant was confirmed \citep{2014Natur.514..198M}. Optical variability has also been observed in several ULXs, such as ULX-4 and ULX-8 (both in M51) and NGC 1313 X-2, suggesting accretion disc-dominated emission in these systems \citep{2009ApJ...690L..39L,2022MNRAS.517.3495A}. Optical counterparts of ULXs are generally faint (m$_{V}$ > 20 mag) and frequently obscured by surrounding stellar populations or interstellar material, which complicates their detection and interpretation. In contrast, infrared (IR) observations, particularly in the NIR range, are less affected by dust extinction and provide valuable insights into the nature of donor stars and the circumstellar environments of ULXs. Several ULXs have reported NIR magnitudes consistent with red supergiants (RSGs) \citealp{2014MNRAS.442.1054H, 2019ApJ...878...71L,2016ApJ...831...88D}, and spectroscopic analyses suggest RSG-like donors in some cases \citep{2016MNRAS.459..771H}. Mid-IR emission is also attributed to either circumstellar dust or jets \citep{2016ApJ...831...88D,2019ApJ...878...71L,2023MNRAS.526.5765A}.
NGC 4631, also known as the Whale Galaxy, is a barred spiral galaxy that exhibits starburst activity. It is located in the constellation Canes Venatici, approximately 7.6 Mpc away (1$^{\prime\prime}$ = 34 pc) \citep{2009ApJ...696..287S}. Several studies have explored the ULX population in NGC 4631, revealing a range of accretion processes. \cite{2007A&A...471L..55C} reported the discovery of a supersoft ULX that exhibits a 4-hour periodic modulation based on \textit{XMM-Newton} data. This supersoft ULX is unlikely to be explained by the standard white dwarf scenario without extreme beaming. It has been suggested that it may instead be a stellar-mass BH accreting at super-Eddington rates, with the modulation resulting from either a companion eclipse or a warped accretion disk.

Furthermore, using archival data from \textit{XMM-Newton}, \textit{Chandra}, and \textit{ROSAT}, \cite{2009ApJ...696..287S} reanalyzed the five brightest X-ray sources in NGC 4631, identifying four (X1, X2, X4, and X5) as ULXs. Among these, X1—previously described as a supersoft source by \cite{2007A&A...471L..55C}—has been proposed to represent an extreme subclass of supersoft sources, powered by transient super-Eddington outbursts driven by non-steady nuclear burning on the surface of a massive white dwarf. The other sources displayed spectral properties consistent with stellar-mass BHs accreting at super-Eddington rates. Moreover, \cite{2025arXiv251104282D} report the discovery of a new transient pulsar ULX in NGC 4631, designated X--8, with a spin period of $9.67~\mathrm{s}$ and a very high spin-up rate, $\dot{P} \sim -9.6 \times 10^{-8}~\mathrm{s/s}$.  \cite{2023ApJ...946...72G} studied the optical counterparts of ULXs in NGC 4631 using a broadband and narrowband imaging campaign with \textit{Canada–France–Hawaii Telescope (CFHT)}/MegaCamm, aiming to detect bubble structures around the X-ray sources and examining their accretion properties. They analyzed the stellar surroundings of the sources via extinction-corrected color–magnitude diagrams and isochrone fitting. Their study reveals a highly asymmetric bubble nebula around X4, displaying distinct morphologies in H$\alpha$ and [O III] emission.

NGC 1097 (D = 16.8 Mpc, 1$^{\prime\prime}$ = 82 pc; \citealt{2011ApJS..192...10L}) is typically classified as a barred spiral (SBb) galaxy. It is also classified as a Seyfert 2 galaxy, which hosts an active galactic nucleus (AGN) powered by a central supermassive BH. Consequently, most studies of this galaxy have focused on constructing the SED of its nucleus using X-ray and multiwavelength observations \citep{2006ApJ...643..652N}. Observations with \textit{Chandra} data further revealed two ULX sources, X2 and X3, in the galaxy, with model-estimated luminosities of $7\times10^{39}$ erg s$^{-1}$ and $3\times10^{39}$ erg s$^{-1}$, respectively \citep{2011ApJS..192...10L}.

\begin{figure*}
\begin{center}
\includegraphics[angle=0,scale=0.35]{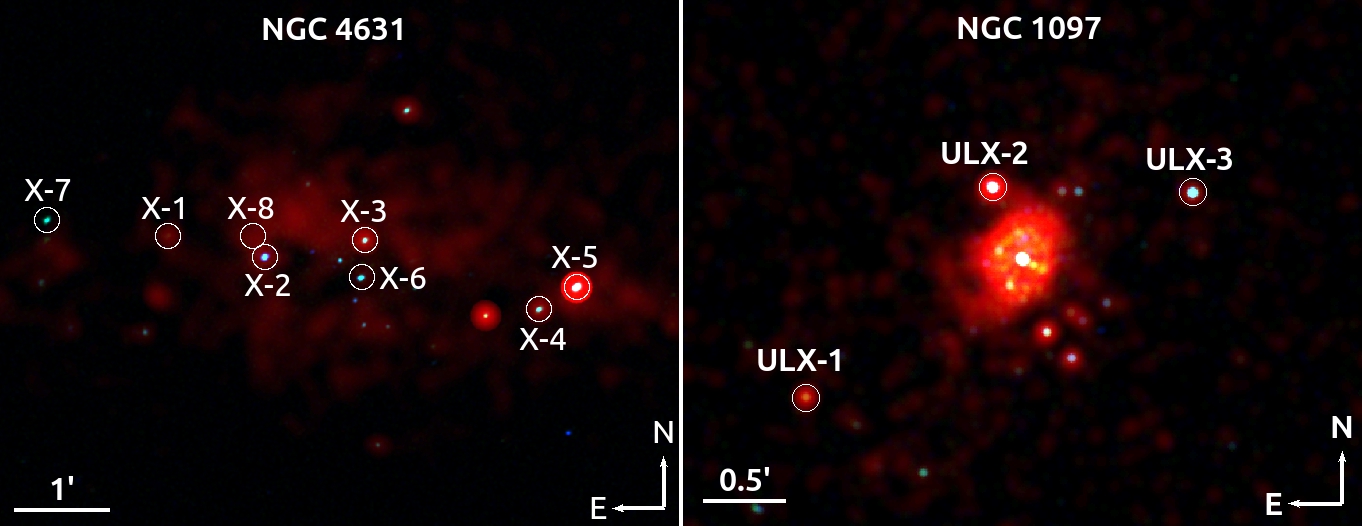}
 \caption{X-ray color-stacked \textit{Chandra} images of the galaxies NGC 4631 (left) and NGC 1097 (right). The energy ranges used for the red, green, and blue colors in both images are 0.5-1.2, 1.2-2, and 2-8 keV, respectively. The positions of the ULXs are indicated with solid white circles.}
\label{F:rgb4631}
\end{center}
\end{figure*}

In this study, we investigate the nature of two newly identified ULX sources in NGC 4631, designated X-6 and X-7, and one ULX source in NGC 1097, designated ULX-3. To achieve this, we used archival data from multiple wavelengths, including X-ray observations from \textit{Chandra}, \textit{XMM-Newton}, and \textit{Swift}/XRT, as well as optical data from the Hubble Space Telescope (HST). The precise positions of the ULX sources studied in both galaxies are shown on the \textit{Chandra} images in Fig. \ref{F:rgb4631}. In addition, we analyzed infrared data from the James Webb Space Telescope (JWST) for the ULX in NGC 1097. The paper is organized as follows: Section \ref{sec:2} provides details of the observations. Section \ref{sec:3} describes the data reduction and analysis of X-ray and optical observations. Section \ref{sec:4} presents the results and discusses the properties of the ULXs. Finally, Section \ref{sec:5} summarizes the main findings of this study.

\section{Observations} \label{sec:2}

The NGC 4631 galaxy was observed with the \textit{Chandra} ACIS-S instrument over 12 observations between 2000 and 2023. Archival observations of the galaxy also include data from \textit{XMM-Newton} taken in 2002, two observations in 2021, and one in 2025. The \textit{HST} observed the galaxy in 2003 and 2014 using the Advanced Camera for Surveys/Wide Field Channel (ACS/WFC). NGC 4631 was also observed 24 times by \textit{Swift}/XRT between 2013 and 2021, with individual exposure times ranging from 128 to 7369 s. In addition, between 8 June 2024 and 9 January 2025, the JWST/Mid-Infrared Instrument (MIRI) observed NGC 4631 using the F1130W, F1500W, F1800W, F560W, and F770W filters. We did not use the \textit{JWST} data  in this study because their spatial resolution (0.11 arcsec/pixel $\sim$ 4 pc/pixel) was insufficient to resolve the counterparts of X-6 and X-7.

The \textit{Chandra} ACIS-S instrument observed NGC 1097 in six separate sessions between 2001 and 2025. The initial observation took place on January 28, 2001, followed by five observations in 2025 (three on 21 May, two on 31 May). The galaxy also has \textit{XMM-Newton} data from an observation conducted on 22 December 2023. In addition, the  \textit{JWST}/NIRCam (Near-Infrared Camera) and MIRI (Mid-Infrared Instrument) instruments observed NGC 1097 with multiple filters between 27 December 2023 and 29 September 2024. The \textit{HST} also observed the galaxy on multiple occasions using different optical filters between 5 June 2004 and 20 August 2025. For \textit{Swift}/XRT, 86 photon counting (PC) mode observations associated with Target IDs 00045597 and 00036582 were carried out between 2007 and 2025. Table \ref{T:obs} provides additional details on the observations of NGC 4631 and NGC 1097. 

\section{X-ray and multiwavelength data reduction and analysis} \label{sec:3}

\subsection{X-ray energy spectra}

We processed the \textit{Chandra} ACIS-S datasets using \textit{Chandra} Interactive Analysis of Observations (CIAO) \citep{2006SPIE.6270E..1VF} and calibration files from the Calibration Database (CALDB) version 4.11, generating level 2 event files using the \textit{chandra\_repro} script available in \textit{CIAO}. Using the {\it wavdetect} tool in \textit{CIAO}, we identified X-ray sources and applied wavelet scales of 2, 4, 8, and 16 pixels with a detection threshold set to 10$^{-6}$. The positions of the sources newly classified as ULXs in this study, based on \textit{Chandra} observations, are: NGC 4631 X-6 at $\alpha=12^\mathrm{h}42^\mathrm{m}06\fs38$, $\delta=+32\degr32\arcmin22\farcs95$; NGC 4631 X-7 at $\alpha=12^\mathrm{h}42^\mathrm{m}22\fs05$, $\delta=+32\degr32\arcmin59\farcs32$; and NGC 1097 ULX-3 at $\alpha=02^\mathrm{h}46^\mathrm{m}14\fs16$, $\delta=-30\degr16\arcmin04\farcs65$.

We derived spectral data and light curves with \textit{specextract} and \textit{dmextract}, respectively. For \textit{Chandra} observations, we extracted the source counts using a circular region with a radius of 4$^{\prime\prime}$. We used two methods to estimate the background: (i) a nearby circular region with a radius of 8$^{\prime\prime}$, free of point sources, and (ii) an annulus centered on the ULX with inner and outer radii of 6$^{\prime\prime}$ and 10$^{\prime\prime}$, respectively. Both approaches yield consistent net count rates within uncertainties. We analyzed \textit{XMM-Newton} European Photon Imaging Camera (EPIC)  observations using the science analysis system ({\scshape sas})\footnote{\url{https://www.cosmos.esa.int/web/xmm-newton/sas}} v22.0. Response files (RMFs and ARFs) for the EPIC-pn and MOS (Metal Oxide Semiconductor detectors) were generated using the \texttt{rmfgen} and \texttt{arfgen} tasks, respectively. The event selection criteria included PATTERN$\le$12 for European Photon Imaging Camera – Metal Oxide Semiconductor (EPIC-MOS) and PATTERN$\le$4 for PN, with \textit{FLAG=0} applied in both cases. We extracted source and background events from circular regions with radii of 15$^{\prime\prime}$ and 30$^{\prime\prime}$, respectively, using the {\it evselect} task. Because NGC 4631 X-6 and NGC 1097 ULX-3 fall into the chip gap of the PN detector on \textit{XMM-Newton}, we used data from the \textit{MOS} detectors only. We used FTOOLS {\it grppha} to group the source energy spectra. For the \textit{XMM-Newton} \textit{MOS} data, we grouped the spectra with at least 15-20 counts per energy bin to enable $\chi^2$ spectral fitting. For \textit{Chandra} observations, the spectra were grouped with 5-10 counts per energy bin, using C-statistics suitable for Poisson-distributed data. We performed all spectral fits using XSPEC \citep{Arnaud1996}, covering 0.5–8 keV (\textit{Chandra}) and 0.3-10 keV (\textit{XMM-Newton}),  to determine the best-fitting models. In addition, due to limited photon statistics in some \textit{Chandra} observations and considering variability timescales, we combined spectra from closely spaced observations to improve the data's  statistical quality and obtain the best spectral fits.

\begin{figure}
 \resizebox{\hsize}{!}{\includegraphics{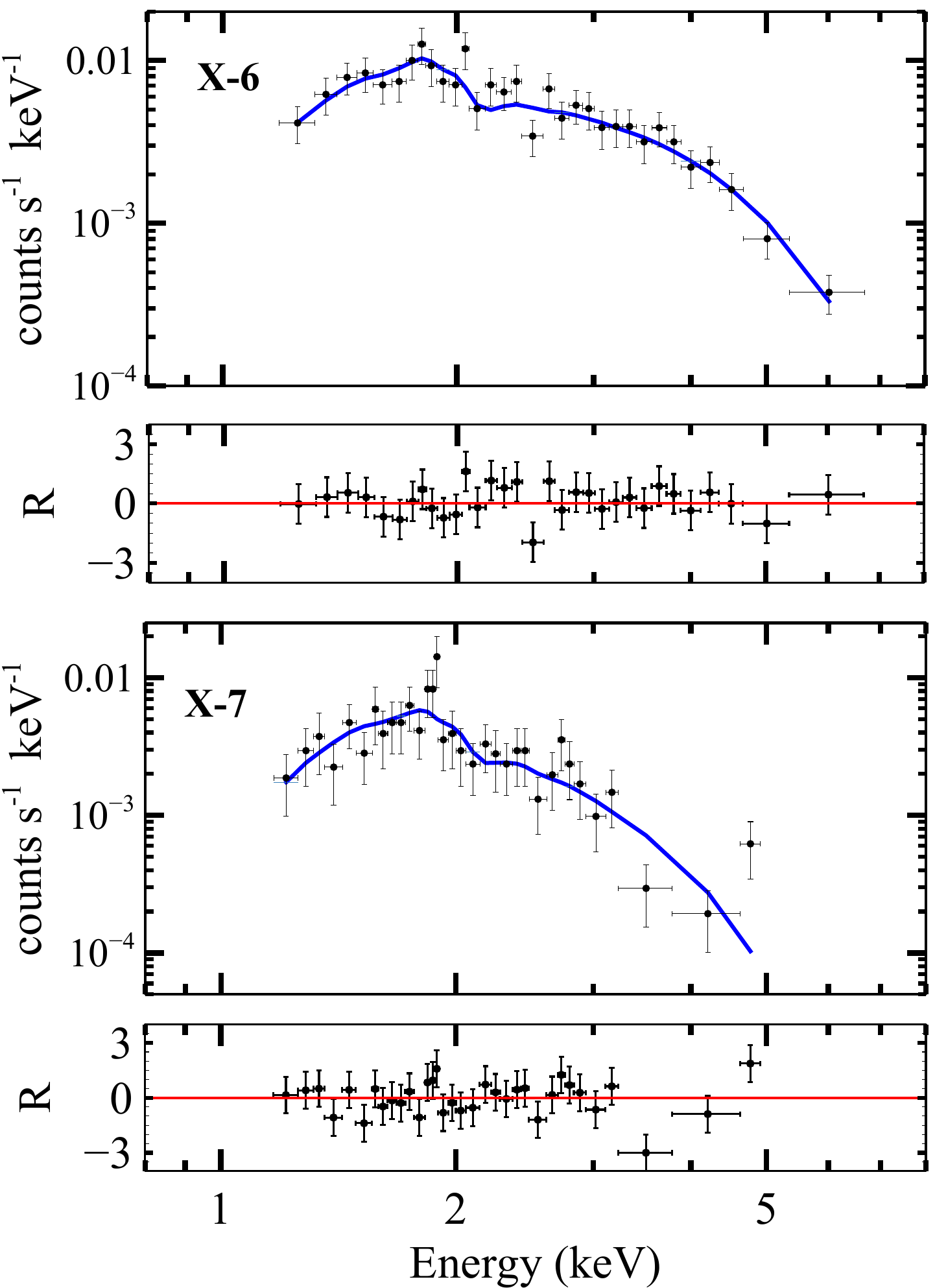}}
 \resizebox{\hsize}{!}{\includegraphics{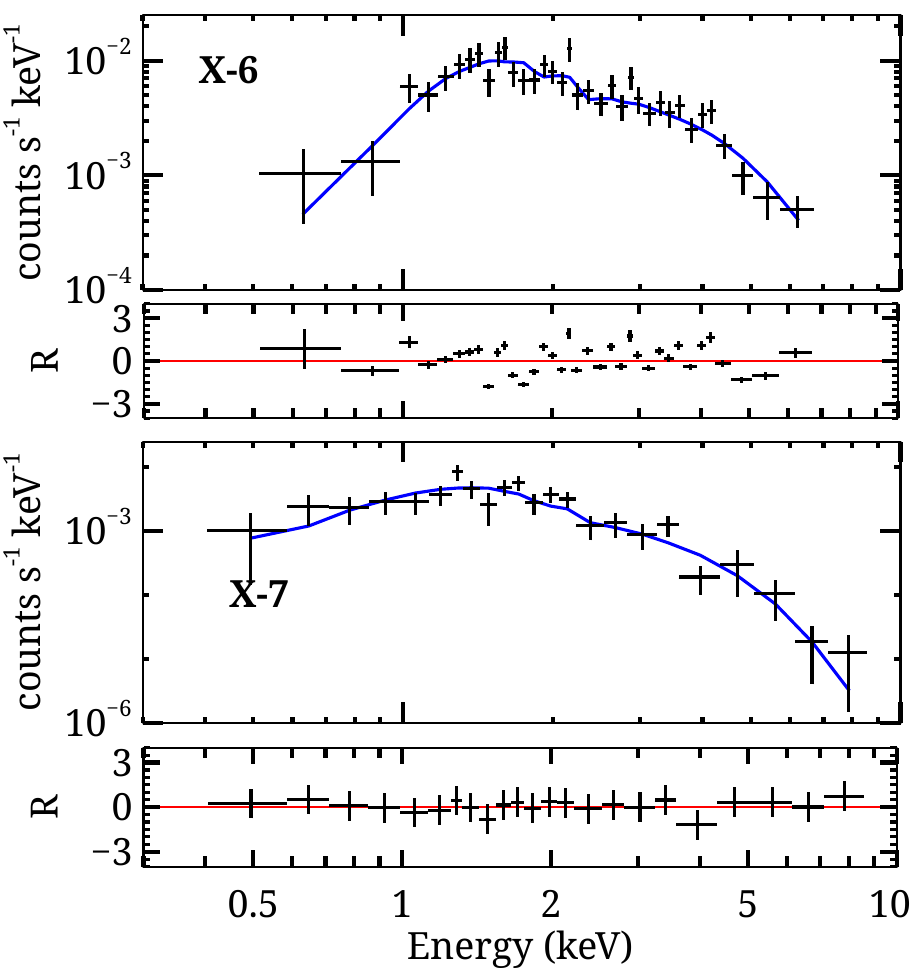}}
 \caption{\textit{Chandra} energy spectra of NGC 4631 ULXs X-6 and X-7 (top two panels) and \textit{XMM-Newton} (MOS) energy spectra (bottom two panels), with residuals shown as $R = (\text{data} - \text{model})/\text{error}$. In all panels, solid blue lines represent the diskbb model.}
 \label{F:spec4631}
\end{figure}

\begin{figure}
 \resizebox{\hsize}{!}{\includegraphics{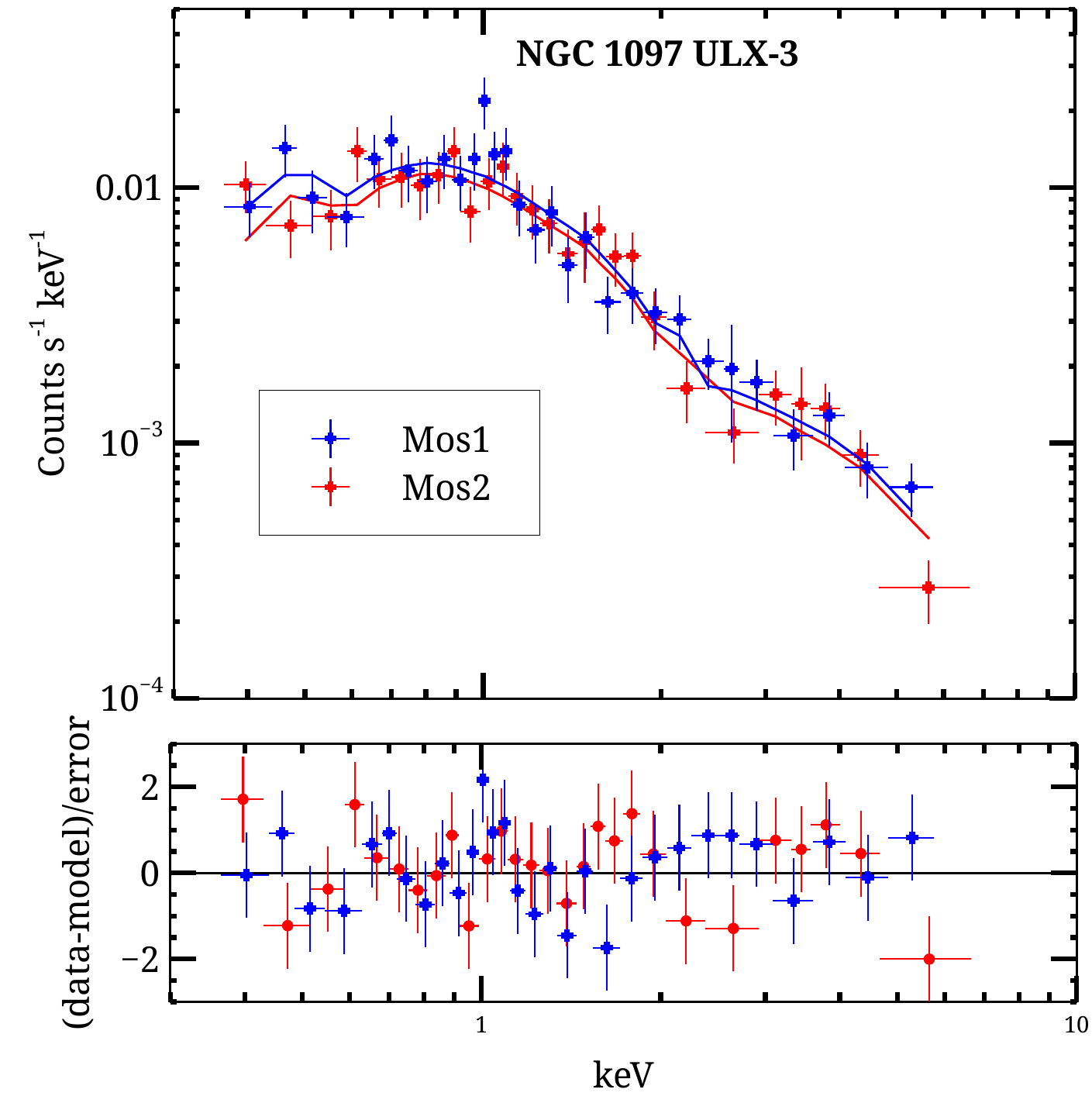}}
 \caption{Energy spectra for NGC 1097 ULX-3 obtained from \textit{XMM-Newton} observations. Blue and red plus symbols represent MOS1 and MOS2 data, respectively. The power-law plus blackbody model fits are shown with green and red lines. All errors are at the 90\% confidence level.}
 \label{F:poapec_ulx3}
\end{figure}

For NGC 4631 ULXs, two-component models did not provide a statistically significant improvement over single-component models. Therefore, we adopted single-component models, either a power-law or a diskbb model combined with two absorption components (tbabs). We used two tbabs components: one fixed at the Galactic hydrogen column density (N$_{\rm H}$ = 0.013 $\times$ 10$^{22}$ cm$^{-2}$) to account for Galactic absorption \citep{1990ARA&A..28..215D}, and the other free to model intrinsic absorption. We initially derived intrinsic $N_{\rm H}$ values from the spectral fits and interpreted the observed variations as more likely statistical than physical. To avoid these fluctuations mimicking flux changes, we repeated the fits with intrinsic $N_{\rm H}$ fixed to the average of the initial values obtained. Intrinsic absorption variability in ULXs remains debated, with reports of both variable $N_{\rm H}$ \citep[e.g.,][]{2009MNRAS.398.1450K} and more stable absorption properties \citep[e.g.,][]{2021A&A...649A.104G}. For NGC 1097 ULX-3, we performed spectral modeling using only the \textit{XMM-Newton} observation, because the \textit{Chandra} observations had insufficient photon counts. with the spectrum best fit by a combined power-law plus bbody model. We provide detailed model parameters for the ULXs, along with the absorbed flux estimated using the \texttt{cflux} convolution model in XSPEC, in Tables \ref{modelx6}, \ref{modelx7}, and \ref{T:spec1079} for X-6, X-7, and ULX-3, respectively. Figures \ref{F:spec4631} and \ref{F:poapec_ulx3} show the corresponding energy spectra.

\begin{table*}
\centering
\caption{X-ray spectral parameters of NGC 4631 X-6 in \textit{Chandra} and XMM–Newton observations.}
\label{modelx6}
\begin{adjustbox}{width=\textwidth}
\begin{tabular}{c c c c c c c c c c }
\hline
Data label & \multicolumn{1}{c}{Net Rate} & \multicolumn{1}{c}{N$_H$}& \multicolumn{1}{c}{$\Gamma$/T$_{in}$} & \multicolumn{1}{c}{$\Gamma^*$/T$_{in}^{*}$} & \multicolumn{1}{c}{$Cstat/dof$} &$F_{x}$ & $L_{x}$ &$F_{x}^{*}$ &$L_{x}^{*}$ \\
 &(10$^{-3}$counts/s) & ($10^{22}$ $cm^{-2}$) &(/keV)& (/keV) & &($10^{-13}$ erg$cm^{-2}$ $s^{-1}$) &($10^{39}$ erg $s^{-1}$) &($10^{-13}$ erg $cm^{-2}$ $s^{-1}$) &($10^{39}$ erg $s^{-1}$) \\
 \\
 \hline
 C1 & >0.08 & & & & & & & &\\
 \multicolumn{10}{c}{\textit{tbabs*tbabs*power-law}} \\ \\
 C2 & 17.64$\pm0.78$ & $1.76^{+0.16}_{-0.14}$ &$2.69^{+0.08}_{-0.07}$ &$2.07^{+0.07}_{-0.07}$& 21.56/28$^b$ &$3.07^{+0.30}_{-0.29}$ &$7.78^{+0.58}_{-0.58}$ & $3.26^{+0.29}_{-0.29}$&$4.12^{+0.31}_{-0.31}$\\
 C3 & 7.93$\pm0.59$ & $0.92^{+0.21}_{-0.19}$ & $2.32^{+0.13}_{-0.12}$ & $2.31^{+0.13}_{-0.12}$ &27.48/29 &$1.48^{+0.29}_{-0.23}$ &$2.16^{+0.27}_{-0.25}$&$1.50^{+0.24}_{-0.21}$ &$2.12^{+0.27}_{-0.25}$\\
 C4 & 9.75$\pm0.73$ & $2.06^{+0.23}_{-0.21}$ & $3.38^{+0.14}_{-0.13}$ & $2.33^{+0.14}_{-0.13}$ &21.99/30 &$1.65^{+0.31}_{-0.25}$ &$8.89^{+1.13}_{-1.04}$&$1.88^{+0.32}_{-0.27}$ &$4.75^{+0.61}_{-0.56}$\\
 C$_{combine}$1 & 9.06$\pm0.47$ &$1.41^{+0.15}_{-0.14}$ &$2.78^{+0.09}_{-0.10}$ & $2.37^{+0.17}_{-0.16}$ &58.47/60&$1.55^{+0.19}_{-0.17}$ &$3.76^{+0.33}_{-0.32}$& $1.64^{+0.18}_{-0.16}$ &$2.07^{+0.19}_{-0.19}$ \\
 C5 &8.25$\pm0.64$ &$1.35^{+0.28}_{-0.26}$ & $1.93^{+0.12}_{-0.12}$ & $1.59^{+0.12}_{-0.12}$ & 35.92/29 &$1.89^{+0.44}_{-0.33}$ &$2.53^{+0.33}_{-0.31}$&$2.00^{+0.37}_{-0.31}$ &$2.07^{+0.28}_{-0.26}$\\
 C6 &9.63$\pm0.81$ &$0.60^{+0.28}_{-0.25}$ & $1.83^{+0.15}_{-0.14}$ & $2.06^{+0.15}_{-0.14}$ &32.27/22& $2.29^{+0.87}_{-0.47}$& $2.29^{+0.33}_{-0.30}$&$2.18^{+0.43}_{-0.35}$&$2.68^{+0.39}_{-0.35}$\\
 C7 & 9.69$\pm0.81$ &$2.10^{+0.32}_{-0.29}$ & $2.71^{+0.14}_{-0.13}$ &$1.83^{+0.14}_{-0.13}$ & 20.52/23&$1.68^{+0.32}_{-0.30}$ & $5.26^{+1.06}_{-0.69}$&$2.14^{+0.41}_{-0.34}$&$2.36^{+0.34}_{-0.31}$\\
 C8 & 4.59$\pm0.39$ &$1.09^{+0.23}_{-0.21}$ & $2.65^{+0.16}_{-0.15}$ & $2.47^{+0.16}_{-0.15}$& 32.46/23&$0.73^{+0.20}_{-0.14}$ &$1.44^{+0.21}_{-0.20}$&$0.77^{+0.15}_{-0.12}$ &$1.21^{+0.18}_{-0.16}$\\
 C$_{combine}$2 & 2.49$\pm0.20$ &$1.67^{+0.20}_{-0.18}$ &$3.65^{+0.17}_{-0.16}$ & $2.80^{+0.17}_{-0.16}$ &22.76/26&$0.39^{+0.08}_{-0.06}$ &$2.47^{+0.34}_{-0.31}$& $0.44^{+0.08}_{-0.06}$ &$0.89^{+0.12}_{-0.11}$ \\
 \hline

\multicolumn{10}{c}{\textit{tbabs*tbabs*diskbb}} \\
 \\
 C2 & & $0.84^{+0.16}_{-0.14}$ & $1.21^{+0.02}_{-0.02}$ & $1.39^{+0.03}_{-0.03}$ & 17.62/28$^b$ &$2.94^{+0.29}_{-0.28}$ &$3.07^{+0.23}_{-0.23}$ &$3.05^{+0.28}_{-0.27}$ &$2.70^{+0.20}_{-0.20}$\\
 C3 & &$0.25^{+0.21}_{-0.19}$ &$1.39^{+0.03}_{-0.03}$ & $1.24^{+0.04}_{-0.04}$&27.22/29&$1.47^{+0.33}_{-0.23}$ & $1.14^{+0.14}_{-0.14}$&$1.40^{+0.23}_{-0.19}$ &$1.25^{+0.16}_{-0.15}$\\
 C4 & &$1.01^{+0.22}_{-0.21}$ &$0.89^{+0.03}_{-0.03}$ & $1.09^{+0.03}_{-0.03}$ & 23.44/30 &$1.55^{+0.27}_{-0.22}$ &$1.93^{+0.25}_{-0.23}$&$1.65^{+0.27}_{-0.23}$ &$1.52^{+0.19}_{-0.18}$\\
 C$_{combine}$1 & &$0.57^{+0.14}_{-0.14}$ &$1.13^{+0.02}_{-0.02}$ & $1.17^{+0.03}_{-0.03}$ &59/60&$1.49^{+0.18}_{-0.16}$ &$1.41^{+0.13}_{-0.12}$& $1.51^{+0.16}_{-0.15}$ &$1.37^{+0.12}_{-0.11}$ \\
 C5 & & $0.77^{+0.28}_{-0.25}$ & $1.69^{+0.06}_{-0.06}$ & $2.01^{+0.08}_{-0.08}$&36.04/29&$1.75^{+0.46}_{-0.31}$ & $1.64^{+0.22}_{-0.20}$&$1.83^{+0.40}_{-0.30}$ &$1.56^{+0.20}_{-0.18}$\\
 C6 & &$0.11^{+0.28}_{-0.11}$ & $1.70^{+0.07}_{-0.07}$&$1.39^{+0.05}_{-0.05}$ &31.76/22&$2.24^{+0.57}_{-0.52}$ & $1.57^{+0.23}_{-0.21}$ &$1.98^{+0.42}_{-0.32}$ &$1.72^{+0.25}_{-0.23}$\\
 C7 & &$1.22^{+0.31}_{-0.29}$ &$1.16^{+0.04}_{-0.04}$ & $1.27^{+0.05}_{-0.05}$ &17.82/23 &$1.76^{+0.35}_{-0.23}$ &$2.05^{+0.29}_{-0.27}$&$1.76^{+0.35}_{-0.28}$ & $1.63^{+0.24}_{-0.22}$\\
 C8 & &$0.45^{+0.23}_{-0.21}$ &$1.04^{+0.04}_{-0.04}$ & $1.02^{+0.04}_{-0.04}$ &30.64/23&$0.67^{+0.19}_{-0.12}$ &$0.63^{+0.09}_{-0.08}$&$0.67^{+0.13}_{-0.10}$ & $0.64^{+0.09}_{-0.08}$\\
 C$_{combine}$2 & &$0.81^{+0.20}_{-0.18}$ &$0.73^{+0.02}_{-0.02}$ & $0.83^{+0.02}_{-0.02}$ &20.85/26&$0.36^{+0.05}_{-0.06}$ &$0.50^{+0.07}_{-0.06}$&$0.38^{+0.06}_{-0.05}$ &$0.40^{+0.05}_{-0.05}$\\
\\
XM1 & >2.74 & & & & & &&&\\
XM2 &21.57 $\pm$ 0.93 &0.71$^{+0.17}_{-0.22}$& 1.40$^{+0.20}_{-0.24}$ & & 34.53/33 & 3.80$^{+0.14}_{-0.16}$ & $2.30^{+0.08}_{-0.10}$&- &-\\ \\
XM3 & >2.11 & & & & & & & &\\ \\
\hline
\end{tabular}
\end{adjustbox}
\vspace{2mm}
\parbox{\textwidth}{%
\footnotesize
{Notes:
The $N_{\mathrm H}$ of one tbabs component was fixed to the line-of-sight value of $0.013 \times 10^{22}\,\mathrm{cm^{-2}}$. The intrinsic $N_{\mathrm H}$ (second tbabs component) was left to free to vary; for the entries marked with (*), the intrinsic $N_{\mathrm H}$ was fixed at average values of $0.90 \times 10^{22}\,\mathrm{cm^{-2}}$ and $0.49 \times 10^{22}\,\mathrm{cm^{-2}}$ for the power-law and diskbb models, respectively.} C$_{combine}$1 and C$_{combine}$2 correspond to combined observations C3--C4 and C9--C12, respectively. The count rate values for C9--C12 are (1.68$\pm$0.38)$\times$10$^{-3}$, (2.07$\pm$0.32)$\times$10$^{-3}$, (2.42$\pm$0.35)$\times$10$^{-3}$, and (2.57$\pm$0.44$)\times$10$^{-3}$, respectively. The superscript b indicates that the corresponding spectral fits were performed using the $\chi^2/dof$ statistic. Absorbed fluxes were calculated in the 0.5–8 keV energy range, while unabsorbed luminosities were derived assuming a distance of 7.6 Mpc. Quoted uncertainties correspond to the 90\% confidence level for each parameter}.
\end{table*}

\begin{table*}[ht!]
\centering
\caption{X-ray spectral parameters of NGC 4631 in the \textit{Chandra} and \textit{XMM–Newton} observations.}
\label{modelx7}
\begin{tabular}{lcccccc}
\hline
Data label & \multicolumn{1}{c}{Net Rate} & \multicolumn{1}{c}{N$_H$}& \multicolumn{1}{c}{$\Gamma$/T$_{in}$} & $Cstat/dof$ &$F_{x}$ &$L_{x}$ \\
 &(10$^{-3}$counts/s) & ($10^{22}$ $cm^{-2}$) &(/keV)& &($10^{-13}$ erg $cm^{-2}$ $s^{-1}$) & ($10^{39}$ erg $s^{-1}$)\\
 \\
\hline
\multicolumn{7}{c}
{\textit{tbabs*tbabs*diskbb}} \\
C1 & >0.20 & & & & & \\
C2 & 6.56$\pm$0.48 & $1.25^{+0.20}_{-0.19}$ & $0.60^{+0.01}_{-0.01}$ & 34.08/30 & $0.88^{+0.13}_{-0.11}$ & $1.78^{+0.22}_{-0.21}$ \\
C3 & 7.72$\pm$0.58 & $0.66^{+0.21}_{-0.19}$ & $0.95^{+0.03}_{-0.03}$ & 14.69/27 & $1.09^{+0.22}_{-0.16}$ & $1.16^{+0.15}_{-0.14}$ \\
C4 & 6.19$\pm$0.58 & $0.17^{+0.20}_{-0.17}$ & $0.87^{+0.03}_{-0.03}$ & 10.10/18 & $0.93^{+0.30}_{-0.22}$ & $0.75^{+0.12}_{-0.11}$ \\
C5 & 8.55$\pm$0.64 & $0.91^{+0.20}_{-0.18}$ & $0.69^{+0.02}_{-0.02}$ & 26.59/28 & $1.01^{+0.17}_{-0.14}$ & $1.61^{+0.21}_{-0.19}$ \\
C6 & 7.75$\pm$0.72 & $0.14^{+0.23}_{-0.14}$ & $0.78^{+0.03}_{-0.03}$ & 12.44/17 & $1.05^{+0.32}_{-0.29}$ & $0.96^{+0.16}_{-0.14}$ \\
C7 & 6.44$\pm$0.66 & $0.92^{+0.27}_{-0.25}$ & $0.55^{+0.02}_{-0.02}$ & 12.38/15 & $0.70^{+0.28}_{-0.15}$ & $1.46^{+0.26}_{-0.24}$ \\
C$_{combine}$ & 7.68$\pm$0.29 & $0.67^{+0.09}_{-0.08}$ & $0.73^{+0.01}_{-0.01}$ & 90.12/99 & $0.95^{+0.07}_{-0.07}$ & $1.20^{+0.08}_{-0.07}$ \\
C8 & >0.49 & & & & & \\
C9 & >0.97 & & & & & \\
C10 & >0.29 & & & & & \\
C11 & >0.23 & & & & & \\
C12 & >0.53 & & & & & \\
\\
XM1 & >0.83 & & & & & \\
XM2 & 8.18$\pm$0.78 & $0.20^{+0.07}_{-0.06}$ & $0.94^{+0.30}_{-0.32}$ & 19.30/20 & $0.82^{+0.04}_{-0.06}$ & $0.50^{+0.06}_{-0.09}$ \\
XM3 & >0.55 & & & & & \\
\hline
\end{tabular}
\begin{flushleft}
Notes: The $N_{\mathrm H}$ of one tbabs component was fixed to the line-of-sight value of $0.013 \times 10^{22}\,\mathrm{cm^{-2}}$, while the intrinsic $N_{\mathrm H}$ (second tbabs component) was left to free to vary. C$_{combine}$ corresponds to combined observations C3–C7. Absorbed fluxes were calculated in the 0.5–8 keV energy range, while unabsorbed luminosities were derived assuming a distance of 7.6 Mpc. Quoted uncertainties correspond to the 90\% confidence level for each parameter.
\end{flushleft}
\end{table*}

\begin{table}
\caption{{Parameters of NGC 1097 ULX-3 from the tbabs*tbabs*(power-law+bbody}) model, using the \textit{XMM-Newton} MOS spectrum.}
\label{T:spec1079}
\begin{tabular}{cccccll}
\hline
Parameters & Units & Values \\
\hline
N$_{\mathrm{H}}$ (a) & $10^{22}$ $\times$ ($\mathrm{cm^{-2}}$) & $0.09^{+0.03}_{-0.02}$\\
${\Gamma}$ & & $1.42^{+0.33}_{-0.28}$ \\
N$_{\mathrm{{\it norm}}}$ & 10$^{-5}$$\times$ (keV$^{-1}$ cm$^{-2}$ s$^{-1}$)& $2.46^{+0.55}_{-0.48}$\\
kT & keV & $0.15^{+0.05}_{-0.04}$ \\
N$_{\mathrm{{\it norm}}}$ & 10$^{-6}$ $\times$ ($L_{39}/D_{10}^2$)& $1.3^{+0.90}_{-0.77}$\\
F$_{\mathrm{X}}$(b) & 10$^{-13}$ $\times$ (erg cm $\mathrm{^{-2}}$ $\mathrm{s^{-1}}$) & $2.05^{+0.09}_{-0.07}$ \\
L$_{\mathrm{X}}$(c) & 10$^{39}$ $\times$ (ergs $\mathrm{s^{-1}}$) & $10.60^{+2.90}_{-2.50}$ \\
$\chi^{2}$/dof & & 56.87/53 \\ 
\hline

\end{tabular}
\\ (a) Intrinsic hydrogen column density ($N_{\mathrm{H}}$) obtained from the spectral fit, with the Galactic absorption fixed at
$N_{\mathrm{H,Gal}} = 0.013 \times 10^{22} \mathrm{cm^{-2}}$.
(b) Absorbed flux in the 0.3–10 keV band. (c) Unabsorbed luminosity in the 0.3–10 keV band, assuming a distance of 16.8 Mpc. All uncertainties are quoted at the 90\% confidence level.\\
\end{table}

\subsection{Short- and long-term X-ray variability}

To investigate the short-term X-ray variability of the three newly identified ULXs, we extracted light curves for all available \textit{Chandra} data in the 0.5–8 keV band and from \textit{XMM-Newton} data in the 0.3–10 keV band. The analysis was performed using time bins of 100, 500, and 1000 seconds. We used the CIAO and SAS tools \texttt{dmextract} and \texttt{evselect}  to background-extract light curves. For the absolute timing analysis, we applied the \texttt{axbary} task to the event file to correct the photon arrival times to the Solar System barycenter, thereby accounting for delays introduced by the orbital motion of the Earth and the \textit{Chandra} spacecraft.

\begin{figure*}
\begin{center}
\includegraphics[angle=0,scale=0.35]{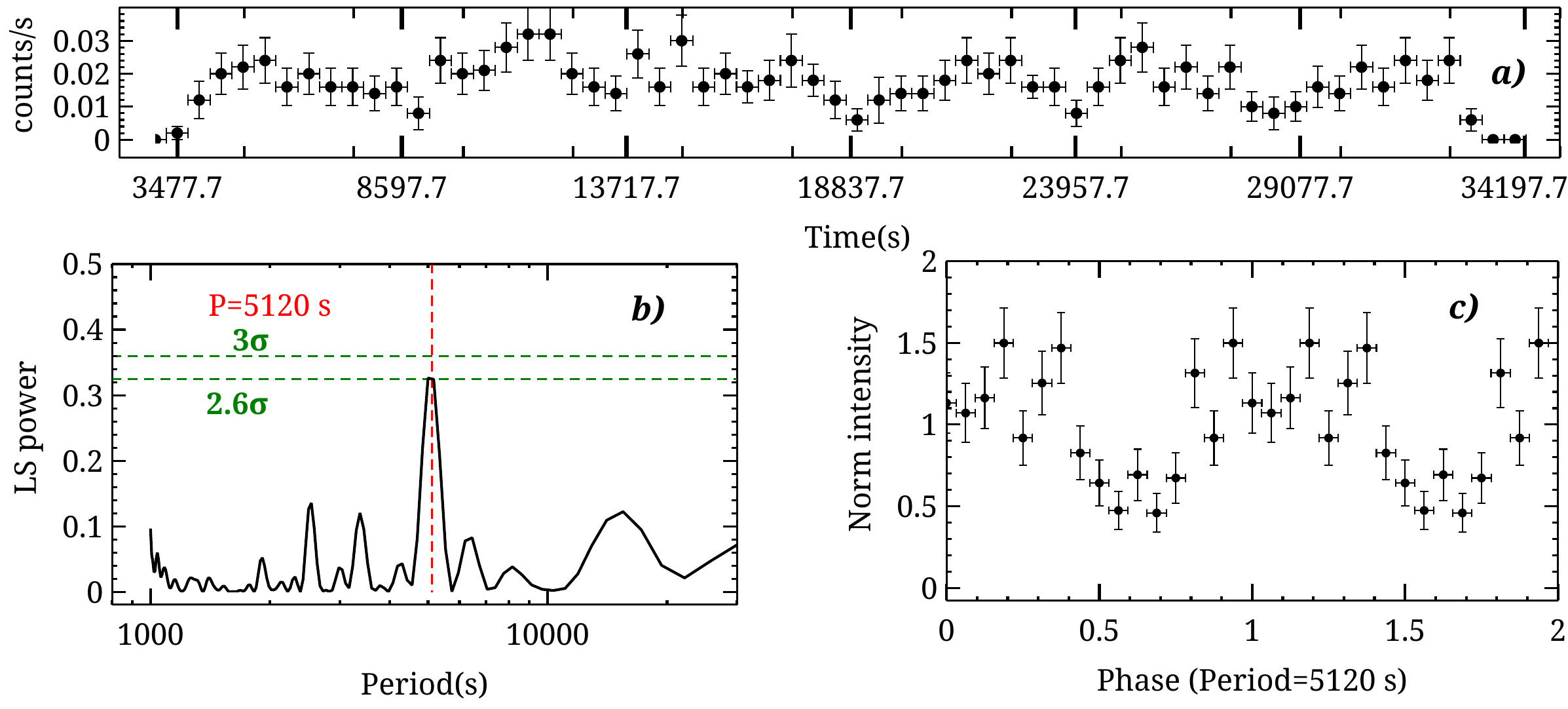}
 \caption{Timing analysis of NGC 4631 ULX X-6 from the C2 observation. 
(a) Background-subtracted light curve in the 0.3--10 keV band with a bin size of 500 s. 
(b) Lomb--Scargle periodogram showing a peak at \textit{P} = 5120 s ($\sim$ 1.42 h) at the 2.6$\sigma$ confidence level. (c) Folded light curve at \textit{P} = 5120 over two cycles, normalized to the mean count rate.}
\label{F:X6LC}
\end{center}
\end{figure*}

We applied the Lomb-Scargle periodogram (LS) and fast Fourier transform (FFT) methods to the light curves to search for periodic signals or variability. In the C2 observation of the NGC 4631 ULX X-6 light curve (Figure \ref{F:X6LC}a), we detected a periodic variation at \textit{P} = 5120 s ($\sim$ 1.42 hr; 2.6-$\sigma$ significance). This signal was identified using both the LS method and with XRONOS from the power density spectrum (Figure \ref{F:X6LC}b). Figure \ref{F:X6LC}c illustrates the folded light curve for this period. To study L$_{X}$ variability, we constructed long-term light curves of X-6 and X-7 from available \textit{Chandra} data. These luminosities were derived from the best-fit reduced C-stat values obtained in the spectral analysis (Tables \ref{modelx6} and \ref{modelx7}). For NGC 1097 ULX-3, we constructed a long-term X-ray light curve using all available archival \textit{Swift}/XRT PC mode observations, along with data from \textit{Chandra} and \textit{XMM-Newton}. We generated the \textit{Swift}/XRT light curve using a custom Python script (\texttt{swifttools}) designed for \textit{Swift}/XRT analysis. The source and background regions were circular, with radii of 30$^{\prime\prime}$ and 60$^{\prime\prime}$, respectively. We converted the resulting count rates to L$_{X}$ (0.3-10 keV) using the spectral model parameters from Table \ref{T:spec1079} and the \texttt{WebPIMMS} tool. For \textit{XMM-Newton}, we derived luminosities from MOS data based on our spectral modeling, while for \textit{Chandra} observations, we utilized the \texttt{srcflux} tool. We show the ULX-3 light curve in Figure \ref{F:lc1097} (left panel).

\subsection{Spectral evolution and state transition}

\begin{figure}
 \resizebox{\hsize}{!}{\includegraphics{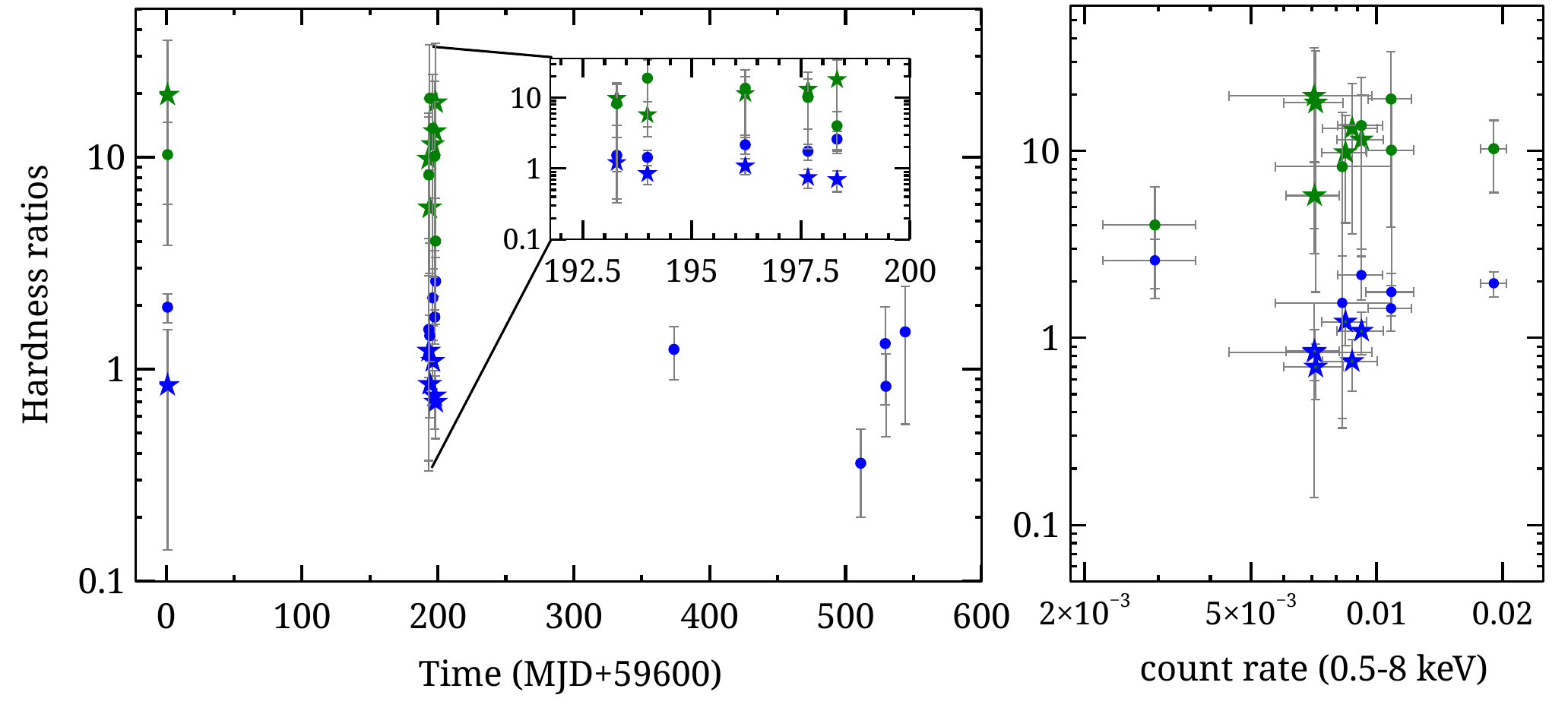}}
 \caption{Left panel: Hardness ratios versus time. Right panel: Hardness–intensity diagram of ULXs X–6 and X–7. X–6 and X–7 are represented by filled circles and stars, respectively. Green and blue colors show the medium/soft and hard/medium ratios, respectively. All error bars represent the 1$\sigma$ level.}
\label{F:hardnessX6X7}
\end{figure}

\begin{figure*}
\begin{center}
\includegraphics[angle=0,scale=0.35]{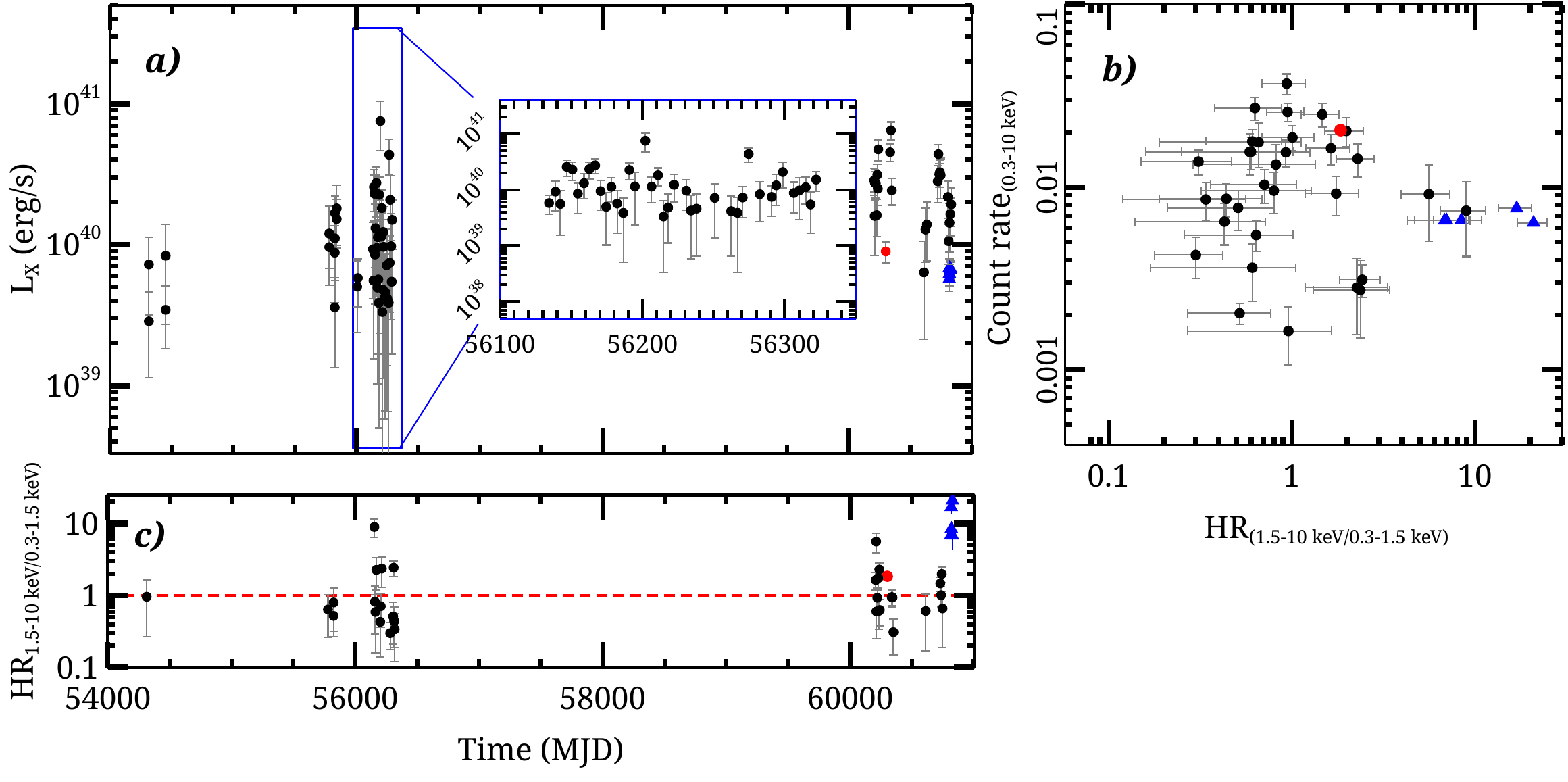}
\caption{Long-term X-ray properties of ULX-3. (a) L$_{X}$ light curve in the 0.3--10 keV band, with the inset showing a zoomed-in view for clarity (MJD 56134-56322). (b) Hardness--intensity diagram, where the count rate in the 0.3--10 keV band is plotted against the hardness ratio HR$_{1.5-10 keV/0.3-1.5 keV}$. (c) Temporal evolution of the hardness ratio HR$_{3.5-10 keV/2-3.5 keV}$ as a function of time (MJD), with the dashed red line marking the reference level. \textit{Swift}/XRT, \textit{XMM-Newton}, and \textit{Chandra} observations are shown with filled black and red circles and blue triangles, respectively.}
\label{F:lc1097}
\end{center}
\end{figure*}

We investigated the spectral evolution of NGC 4631 X-6 and X-7 using \textit{Chandra} observations, excluding non-detections. Soft (S), medium (M), and hard (H) bands were defined as 0.3–1.2 keV, 1.2–2 keV, and 2–8 keV, respectively, and the M/S and H/M hardness ratios (HR) were calculated. The temporal evolution of the hardness ratios is shown in the left panel of Figure \ref{F:hardnessX6X7}, and the hardness--intensity diagram is shown in the right panel. When a band had insufficient counts, we adopted 3$\sigma$ upper limits. To investigate potential spectral state transitions of NGC 1097 ULX-3, we constructed a hardness–intensity diagram (right panel of Figure \ref{F:lc1097}), using count rates in the 0.3--1.5 keV (S) and 1.5--10 keV (H) bands. To account for differences in instrumental responses, we normalized the count rates from \textit{Chandra} and \textit{XMM-Newton} observations to those from \textit{Swift}/XRT using WebPIMMS.

\subsection{Optical and infrared photometry }

For all drizzled HST/Wide Field Camera 3/Ultraviolet and Visible channel (WFC3/UVIS) and HST/Advanced Camera for Surveys/Wide Field Channel (ACS/WFC) images, we identified point-like sources with \texttt{daofind} task, and their fluxes were measured with aperture photometry employing the  DAOPHOT package \citep{1987PASP...99..191S} within {\scshape IRAF}\footnote{\url{https://iraf-community.github.io/}} (Image Reduction and Analysis Facility). For the \textit{JWST}/NIRCam observations, background estimation, source detection, and photometry were performed following the methodology of \cite{2023MNRAS.526.5765A}. Instrumental magnitudes were converted into Vega magnitudes using zero points (ZP) provided for WFC3/UVIS and ACS/WFC taken from \cite{2022AJ....164...32C}, and the ZP calculator\footnote{\url{https://acszeropoints.stsci.edu/}}. Sources were required to exceed a 3$\sigma$ significance above the local background. We then measured photometry using circular apertures of 3-pixel radius, with the background contribution subtracted using an annulus located nine pixels from the centroid.

\begin{figure}
 \resizebox{\hsize}{!}{\includegraphics{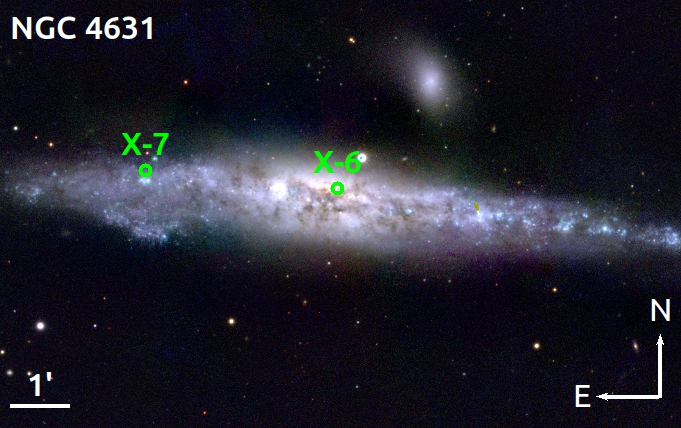}}
 \resizebox{\hsize}{!}{\includegraphics{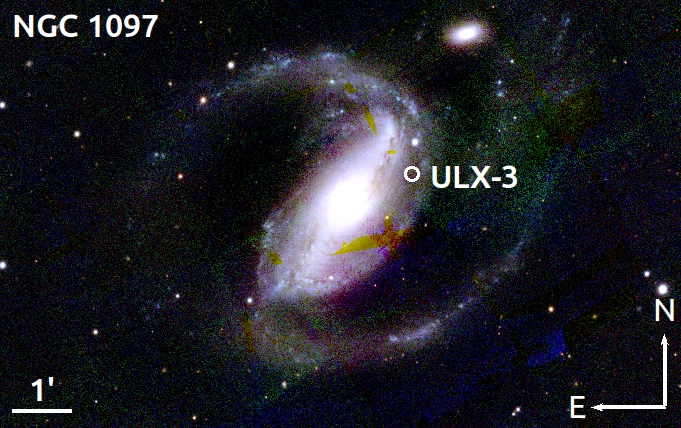}}
 \caption{Visualization of new ULXs in NGC 4631 (left) and NGC 1097 (right), on Pan-STARRS RGB (i, r, g) composite images. The positions of the ULXs are indicated with green and white circles.}
 \label{F:COUNTERPART}
\end{figure}

\begin{figure}
 \resizebox{\hsize}{!}{\includegraphics{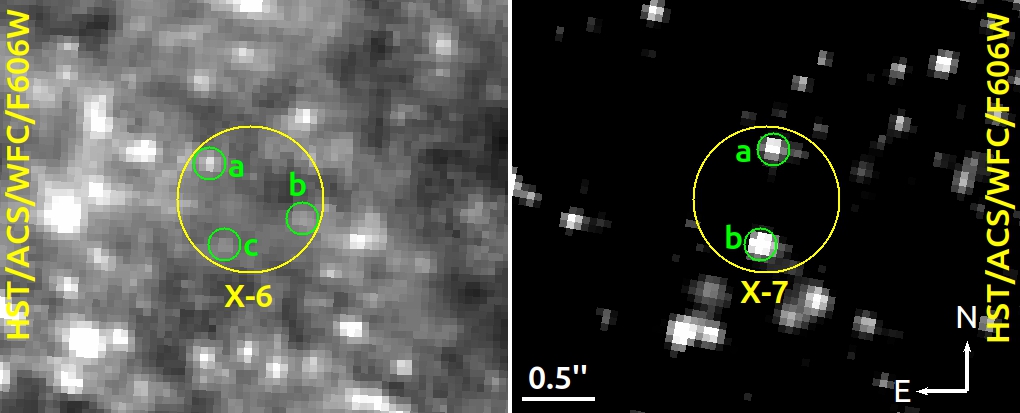}}
 \resizebox{\hsize}{!}{\includegraphics{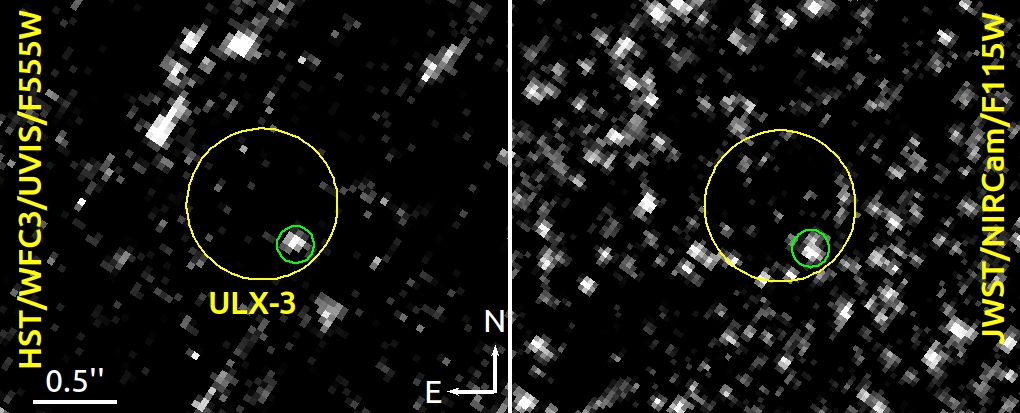}}
 \caption{Upper panels: Optical counterparts of ULXs X-6 and X-7 in NGC 4631 (upper panels). The solid yellow circles have a radius of 0.5$^{\prime\prime}$. The small green circles represent the candidate counterparts. Bottom panels: Position of the optical counterpart (green circle) of ULX-3 on \textit{HST} (left) and \textit{JWST} (right) images. The solid yellow circles show the astrometric error radius of 0.45$^{\prime\prime}$}.
\label{F:COUNTERPARTs}
\end{figure}

Using the astrometric corrections reported by \cite{2023ApJ...946...72G}, we identified three optical counterparts (X6a-Xc) for X-6 and two (X7a and X7b) for X-7 in the F606W and F814W bands. For NGC 1097 ULX-3, we searched the available catalogs (2MASS and Gaia DR3) for optical and NIR counterparts. Aside from the galaxy nucleus, we did not identify any reference sources with XRBs in these catalogs; therefore, the nucleus was used as the reference source for astrometric corrections. We corrected the relative astrometry between \textit{Chandra} (CH7) and \textit{HST} (F555W), \textit{Chandra} and \textit{JWST} (F115W), and \textit{HST} and JWST, obtaining $95\%$ confidence error radii of $0\farcs45$, $0\farcs27$, and $0\farcs10$, respectively. Based on these results, we identified a unique optical and NIR counterpart within the astrometric error radius by applying a 3$\sigma$ detection threshold over the background. Figure \ref{F:COUNTERPART} shows the positions of the ULXs in Pan-STARRS RGB composite images (i, r, g bands), illustrating their distribution across the galaxy. In the NIRCam images (Figure \ref{F:COUNTERPARTs}), several faint NIR sources are present around the counterpart, but were not detected above the 1.5$\sigma$ level. Therefore, we adopted the optical counterpart as the corresponding NIR counterpart of ULX-3.  Figure \ref{F:COUNTERPARTs} shows the positions of all three ULX counterparts. Tables \ref{T:optical_NGC4631} and \ref{T:optical_ULX3} list the photometric results for the optical and NIR counterparts of NGC 4631 ULX X-6 and X-7, and NGC 109 ULX-3.

\begin{table}[h!]
\raggedright
\caption{Vega magnitudes of ULXs X-6 and X-7 optical counterparts in NGC 4631.}
\label{T:optical_NGC4631}
\begin{tabular}{ccc}
\hline
Source & Filter & Vega magnitude \\
\hline
X6a & F606W HST/WFC3 & 25.50 $\pm$ 0.06 \\
X6a & F814W HST/WFC3 & 26.35 $\pm$ 0.08 \\
X6b & F606W HST/WFC3 & 26.45 $\pm$ 0.08 \\
X6b & F814W HST/WFC3 & 25.48 $\pm$ 0.08 \\
X6c & F606W HST/WFC3 & 25.86 $\pm$ 0.04 \\
X6c & F814W HST/WFC3 & 25.11 $\pm$ 0.10 \\
X7a & F606W HST/WFC3 & 24.53 $\pm$ 0.04 \\
X7a & F814W HST/WFC3 & 24.20 $\pm$ 0.06 \\
X7b & F606W HST/WFC3 & 24.31 $\pm$ 0.04 \\
X7b & F814W HST/WFC3 & 24.37 $\pm$ 0.06 \\
\hline
\end{tabular}

\end{table}

\begin{table}[h!]
\raggedright
\caption{Vega magnitudes of optical and infrared counterparts of ULX-3.}
\label{T:optical_ULX3}
\begin{tabular}{cccc}
\hline
Filter & Instrument & Date & Vega magnitude \\
\hline
F275W & HST/WFC3 & 2020-08-14&26.29 $\pm$ 0.47 \\
F336W & HST/WFC3 & 2014-02-20&26.55 $\pm$ 0.38 \\
F438W & HST/WFC3 & 2014-02-20&>27.60 \\
F555W & HST/WFC3 &2020-08-14 &>27.40 \\
F814W & HST/ACS/WFC & 2004-06-05 &>26.50\\
F814W & HST/WFC3 & 2014-02-20& >26.00 \\
F814W & HST/ACS/WFC & 2019-05-28&25.74 $\pm$ 0.26 \\
F115W & \textit{JWST}/NIRCam & 2023-12-27&22.97 $\pm$ 0.08 \\
F115W & \textit{JWST}/NIRCam & 2024-09-29&22.55 $\pm$ 0.04 \\
F200W & \textit{JWST}/NIRCam & 2024-09-29&22.40 $\pm$ 0.04 \\
F277W & \textit{JWST}/NIRCam &2024-09-29 &22.25 $\pm$ 0.09 \\
F300M & \textit{JWST}/NIRCam & 2023-12-27&22.25 $\pm$ 0.11 \\
F335M & \textit{JWST}/NIRCam & 2023-12-27&22.25 $\pm$ 0.09 \\
F335M & \textit{JWST}/NIRCam & 2024-09-29&22.45 $\pm$ 0.08 \\
F360M & \textit{JWST}/NIRCam & 2024-09-29&22.40 $\pm$ 0.08 \\
F444W & \textit{JWST}/NIRCam & 2024-09-29&22.75 $\pm$ 0.09 \\
\hline
\end{tabular}

\end{table}

\begin{figure}
\resizebox{\hsize}{!}{\includegraphics{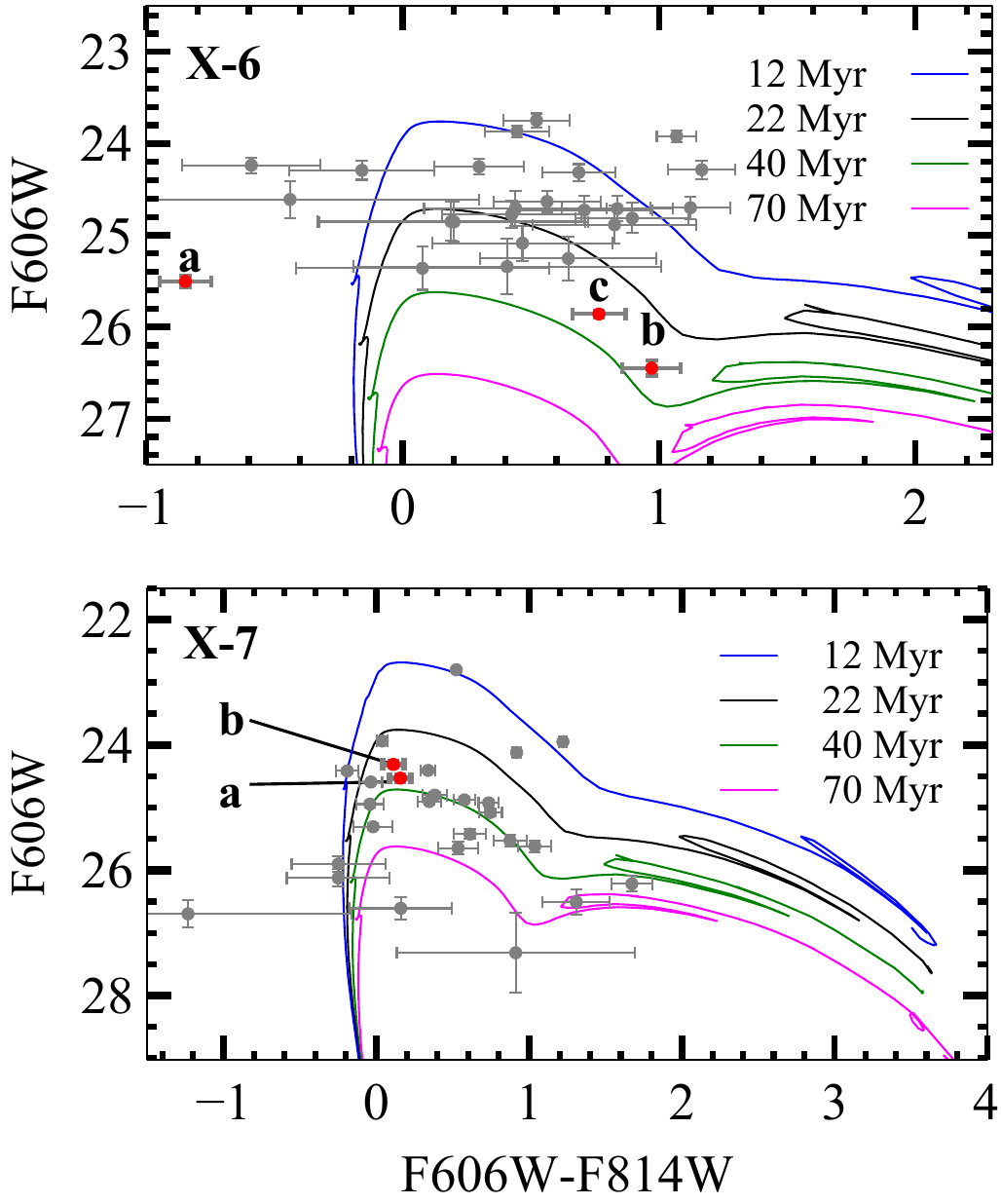}}
\caption{Color-magnitude diagrams of NGC 4631 X-6 and X-7. The filled red and gray circles represent candidate counterparts and field stars, respectively.}
\label{F:cmds}
\end{figure}

\begin{figure}
\resizebox{\hsize}{!}{\includegraphics{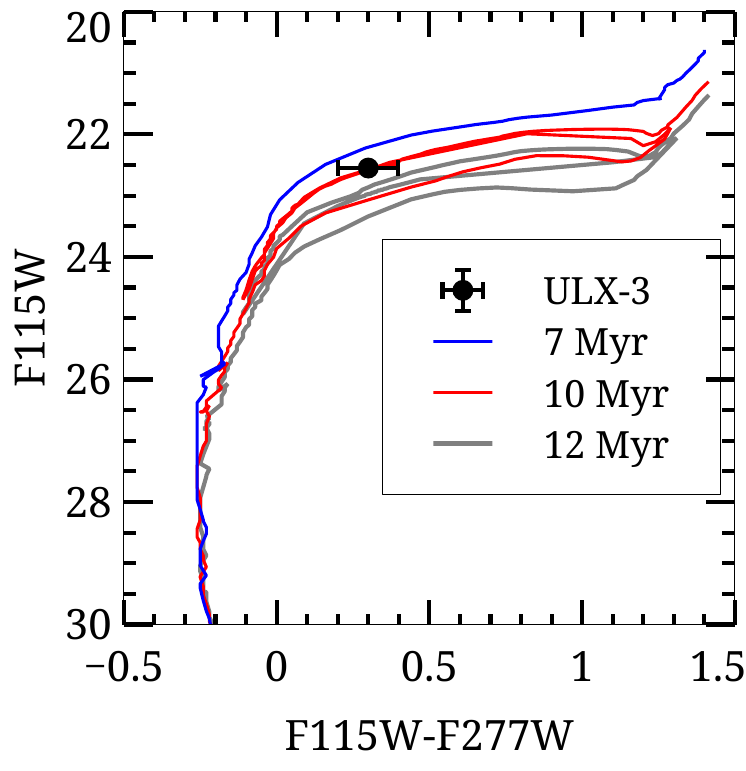}}
\caption{Color-magnitude diagram of NGC 1097 ULX-3.}
\label{F:cmdsx3}
\end{figure}

Color–magnitude diagrams (CMDs) provide insights into the ages of the stellar populations surrounding ULXs, which may help constrain the nature of their donor stars. Isochrone models from the PAdova TRieste Stellar Evolution Code (PARSEC) \citep{2012MNRAS.427..127B} stellar evolution code are available for various telescope filters, including those of \textit{HST}/ACS. We derived the upper limit extinction, A$_{V}$, using the method of \cite{2023ApJ...946...72G} and used this value to correct the isochrones. Using the average N$_{H}$ values from Tables \ref{modelx6} and \ref{modelx7}, we derived E(F606W-F814W) values of 2.90 mag and 1.72 mag for X-6 and X-7, respectively. Figure \ref{F:cmds} shows the CMDs for X-6 and X-7, plotted as F606W versus F606W-F814W, with isochrones and field sources located within 2$^{\prime\prime}$ radius circles centered on the ULX positions, alongside the optical counterparts. For ULX-3, we only constructed an NIR CMD due to the limited optical data. The \textit{HST}/F814W photometry shows variability likely originating from the accretion disk; nonsimultaneous filters prevent reliable color measurements. The source is extremely faint in the UVIS bands and F438W detection is consistent with an upper limit, making an optical CMD infeasible. In contrast, the \textit{JWST}/NIRCam data show no variability; therefore, we were able to construct a single-epoch NIR CMD. The NIR CMD was constructed using F115W and F277W photometry, adopting a distance of 16.8 Mpc and an upper limit extinction of 0.06 mag (Figure \ref{F:cmdsx3}).

\begin{figure}
 \resizebox{\hsize}{!}{\includegraphics{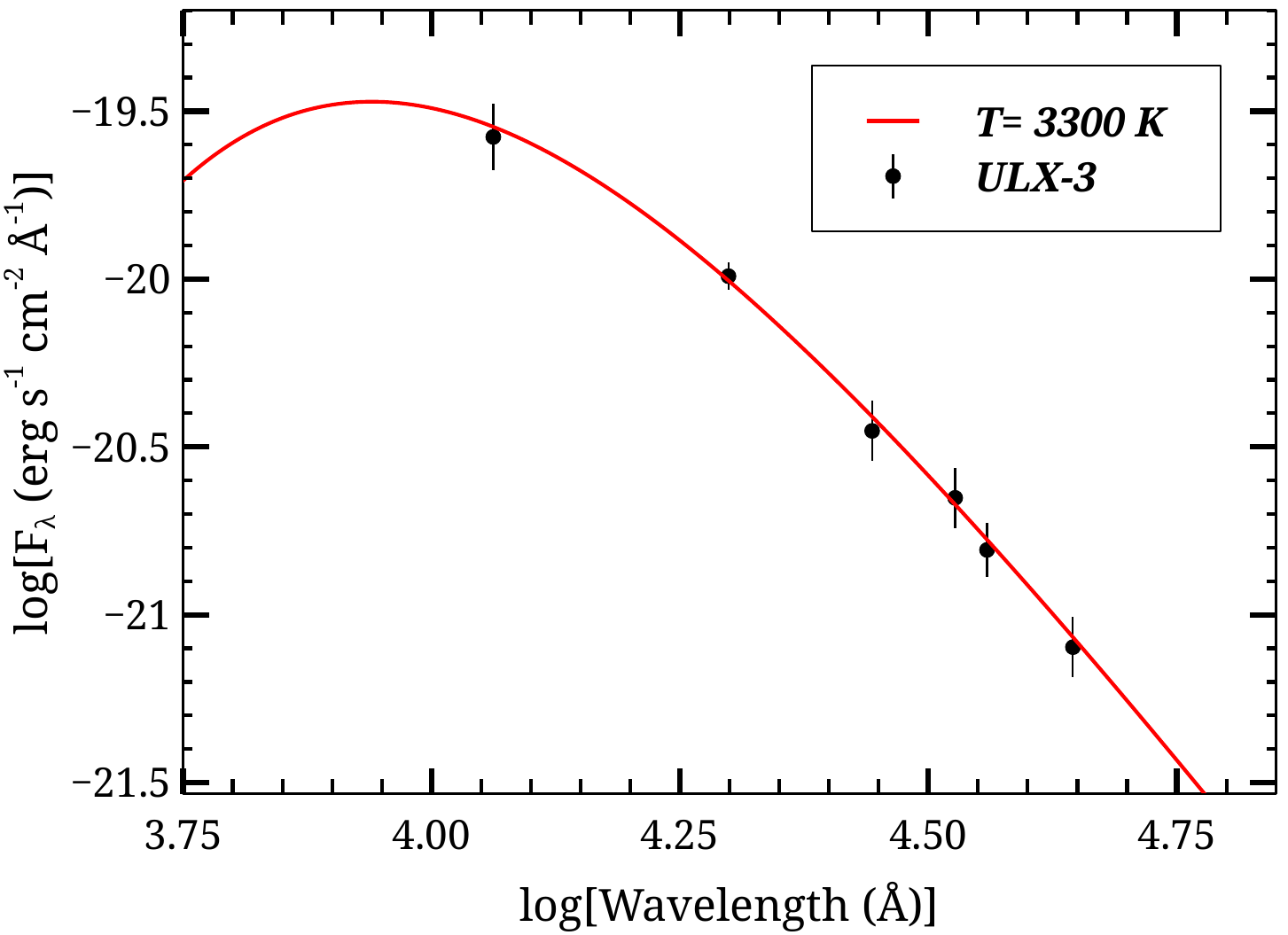}}
 \caption{Near-IR SED of NGC 1097 ULX-3. The solid red line represents a blackbody model with a temperature of 3300 K. The SED is constructed from simultaneous NIRCam observations.}
 \label{F:sedulx3}
\end{figure}

For ULX-3, we converted the observed magnitudes  into fluxes using the corresponding zeropoints and adopted the pivot wavelength of each \textit{JWST}/NIRCam filter. The resulting SED is well fit by a blackbody model with a best-fit temperature of \textit{T} = 3300 $\pm$ 140 \textit{K} and a reduced chi--square of $\chi^{2}/{\rm dof}$ = 0.95 for four degrees of freedom (Figure \ref{F:sedulx3}).

\section{Results and discussion}\label{sec:4}

\subsection{NGC 4631 X-6 and X-7}

\subsubsection{Spectral modeling and variability}

In NGC 4631, we identified two new transient ULX candidates, X-6 and X-7, from observations conducted between 2021 and 2023. Across 13 \textit{Chandra} observations (C1–C13) and three \textit{XMM-Newton} observations (XM1–XM3), we did not detect X-6 in C1, XM1, or XM3. We detected X-7 source only in XM2 and in five \textit{Chandra} observations (C2–C7). We also examined \textit{Swift}/XRT PC observations using the \textit{Swift-XRT} data products tool\footnote{\url{https://www.swift.ac.uk/user_objects/}} to investigate the visibility of X-6 and X-7.

For X-6, both the power-law and diskbb models yield $N_{\mathrm{H}}$ values that vary by nearly an order of magnitude in \textit{Chandra} observations when intrinsic absorption is left free. The $N_{\mathrm{H}}$ variability may reflect genuine changes in the local environment, but could also result from statistical uncertainties or spectral modeling limitations. We therefore fixed $N_{\mathrm{H}}$ to its average value and repeated the spectral fits (Table \ref{modelx6}). After applying this procedure, we obtained notable improvements in the reduced $\chi^{2}$ values for both the diskbb and power-law models. Applying simple phenomenological models to the X-6 spectra, we found that the two models could not be distinguished in the \textit{Chandra} datasets, likely because of limited data quality. The power-law photon index ($\Gamma$) ranges between 1.6 and 2.8. For Galactic BH binaries (BHBs), $\Gamma$ $\sim$1.4–2.4 typically corresponds to the hard state, characterized by Comptonization of disk photons in a hot, optically thin corona \citep[e.g.,][]{2006ARA&A..44...49R,2007A&ARv..15....1D}. For X-6, $\Gamma$ remains <2.4 in nearly all \textit{Chandra} observations (except C8), indicating that the source resides predominantly in the hard state. Based on the power-law model, X-6 exhibits L$_{X}$ = $(1-4)\times10^{39}$ erg s$^{-1}$ in the 0.5–8 keV band.

Furthermore, X-6 is well described by the diskbb model, with the inner disk temperature ($T_{\rm{in}}$) varying between 0.8 and 2 keV. Using the diskbb spectral model, X-6 exhibits L$_{X}$ = $(0.6-2.7)\times10^{39}$ erg s$^{-1}$ in the 0.5–8 keV band. For X-7, among all the single and multi-component models tested in XSPEC, only the diskbb model provides acceptable fits, yielding inner disk temperatures of $kT_{\mathrm{in}} \sim 0.6$–$1$ keV for roughly half of the dataset. For the C2–C7 observations, X-7 exhibits L$_{X}$ = $(1-2)\times10^{39}$ erg s$^{-1}$ in the 0.5–8 keV band, based on the diskbb spectral model. The temperatures indicate that X-ray emission is dominated by accretion disk processes, consistent with Galactic XRBs in the thermal state at high accretion rates. Such disk-dominated spectral behavior has been reported in several ULX studies, with thermal components consistent with super-Eddington accretion or slim disk models \citep[e.g.,][]{2009MNRAS.397.1836G,2016MNRAS.460.4417L}. To search for spectral state transitions, we constructed hardness ratios as a function of time and hardness–intensity diagrams (Figure \ref{F:hardnessX6X7}). Source X-6 predominantly occupies the hard state, with episodes indicating transitions from hard to intermediate and back to hard, whereas source X-7 resides mostly in the intermediate state.

The X-6 and X-7 ULXs do not show significant variability on short timescales of a few days.
We detected a $\sim$5120 s, low-amplitude,  low-significance (2.6-$\sigma$) periodic modulation for NGC 4631 X-6 only in the C2 observation. Hour-scale (4–7 ks) quasi-periodic behavior has been reported in some ULXs (e.g., the M74 ULX; \citealt{2005ApJ...621L..17L}), showing that this timescale is physically plausible for accretion-flow variability. In contrast, orbital periods in ULXs are typically days, except for a few systems with Wolf-Rayet donors \citep[e.g.,][]{2023ApJ...954...46L}, which show hour-scale periods. Because the period falls below conventional significance thresholds, deeper and higher signal-to-noise observations are needed to constrain the origin of this periodic feature.

\subsubsection{$L_{\rm X} - T_{\rm in}$ relation}

\begin{figure}
 \resizebox{\hsize}{!}{\includegraphics{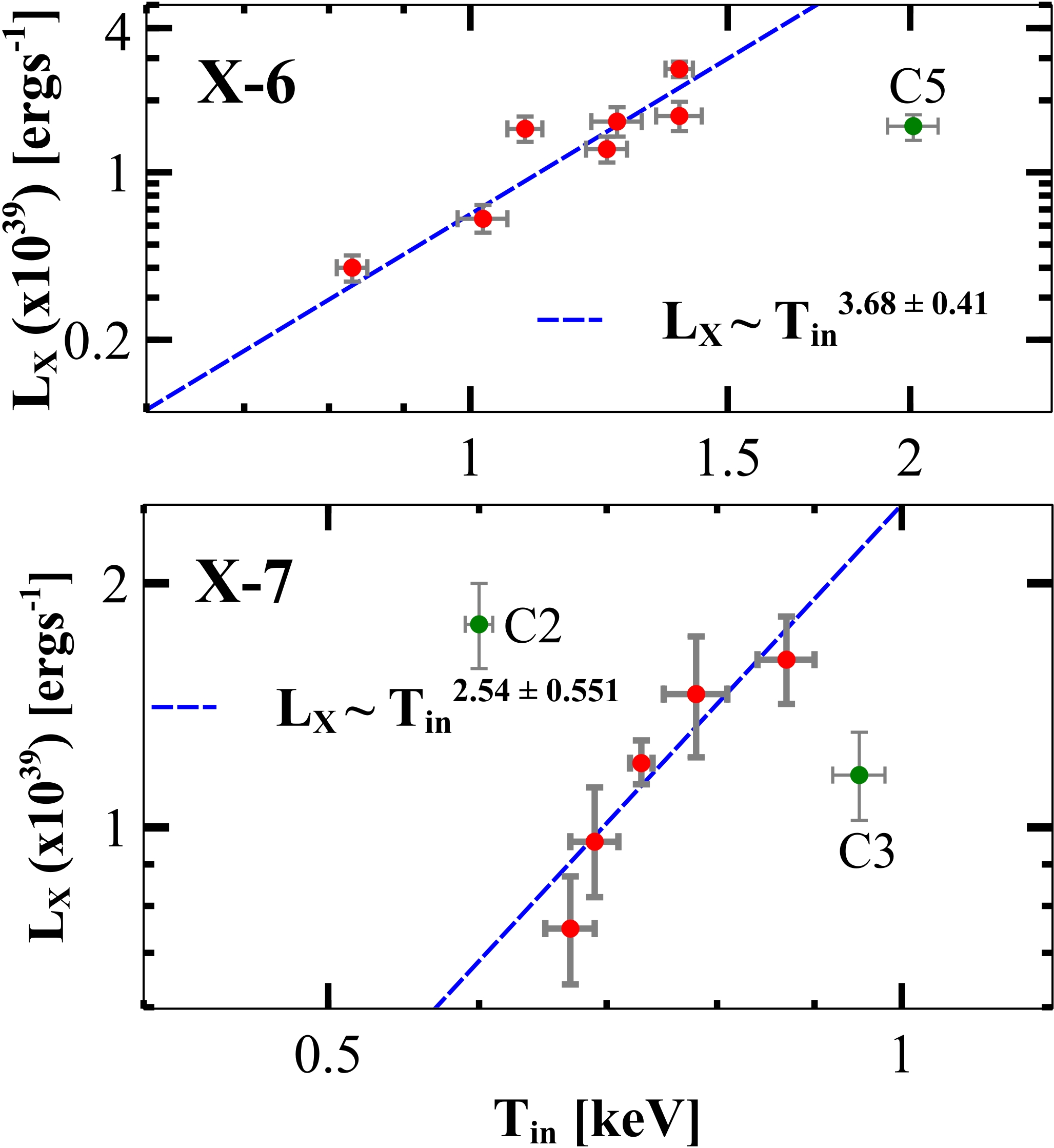}}
 \caption{Luminosity (L$_{X}$) versus inner disk temperature (T$_{in}$) diagrams of X-6 (top) and X-7 (bottom) inferred from the diskbb model. The data points use T$_{\mathrm{in}}^*$ and L$_{X}^*$ and T$_{\mathrm{in}}$ and L$_{X}$ values from Tables \ref{modelx6} and \ref{modelx7}.}
 \label{F:LxTin}
\end{figure}

We investigated the relationship between disk luminosity ($L_{\rm disc}$) and the color temperature of the accretion disk ($T_{\rm in}$) for X-6 and X-7. In Galactic BHBs, the X-ray emission in the thermal dominant state is generally produced by a geometrically thin, optically thick disk, which is well described by the standard accretion disk model \citep{1973A&A....24..337S}. Within this framework, $L_{\rm disc}$ and $T_{\rm in}$ are expected to follow the relation $L_{\rm disc} \propto T_{\rm in}^{4}$ if the inner disk radius remains constant. Testing this correlation probes accretion geometry and physical conditions, while deviations from the expected trend can reveal structural changes in the disk, such as truncation, strong Comptonization, or transitions away from the thin-disk regime, particularly at high luminosities \citep{2004MNRAS.353..980K,2007A&ARv..15....1D}. Figure \ref{F:LxTin} shows the relationship between $L_{\rm X}$ and $T_{\rm in}$ for X-6 using diskbb spectral parameters (Tables~\ref{modelx6} and \ref{modelx7}).
The data reveal a strong positive correlation of the form $L_X \propto T^{\alpha}$, with a best-fit slope $\alpha = 3.68 \pm 0.41$ (dashed line), consistent within the uncertainties with the canonical $L \propto T^4$ relation predicted by the standard accretion disk model \citep{1973A&A....24..337S}. This relation suggests that X-6 is likely in a disk-dominated accretion regime with a relatively stable inner disk radius. The outlier, denoted C5, significantly deviates from the observed trend, lying at a higher temperature but lower luminosity than expected. This may indicate a temporary transition to a different spectral state, obscuration, or a change in viewing geometry. For example, \citet{2007ApJ...660L.113F} found that NGC 1313 X-2 exhibited a relation consistent with $L \propto T^4$ during certain states, indicating a constant inner disk radius. Transient ULXs, such as XMMU J004243.6+412519 in M31, have also shown L–T tracks broadly consistent with the expected $T^{4}$ relation. \citep{2013Natur.493..187M}. Overall, the $L_{\rm X} - T_{\rm in}$ trend observed in X-6 supports a stellar-mass compact object accreting in a high state, rather than an intermediate-mass BH scenario, which typically exhibits cooler disks at high luminosities \citep{2003MNRAS.345.1057M}.

In addition, the diskbb normalization parameter provides an estimate of the apparent inner disk radius: \( N = \left( \frac{R_{\mathrm{in}}/\mathrm{km}}{D_{10}} \right)^{2} \cos\theta \), where \(N\) is the normalization, \(R_{\mathrm{in,obs}}\) is the observed inner disk radius in km, and $D_{10}$ is the distance to the source in units of 10 kpc. Using \(N_{\min} = 1.6 \times 10^{-3}\) and \(N_{\max} = 25 \times 10^{-3}\), the observed inner disk radii are 28 and 110 km. To account for spectral hardening and boundary condition effects, we applied a correction factor of \(\kappa^2 \xi = 1.19\) \citep{1995ApJ...445..780S,1998PASJ...50..667K}. Therefore, the corrected values of R$_{in,min}$ and R$_{in,max}$ are 30 and 130 km, respectively. Assuming that the inner disk radius corresponds to the innermost stable circular orbit (ISCO) of a Schwarzschild BH, $R_{\rm in} \approx R_{\rm ISCO} = 6\,GM/c^2$, the mass is \( \frac{M}{M_{\odot}} = \frac{R_{\mathrm{in}}}{8.86\,\mathrm{km}} \). The estimated mass range is $\sim$5--15\,$M_{\odot}$, increasing to $\sim$5--20\,$M_{\odot}$ for an assumed disk inclination of 60$^\circ$. These values are consistent with a stellar-mass BH.

We calculated the accretion rate to assess the accretion regime of ULXs following the method described in Section 4.4 of \citet{2023ApJ...949...78G}. For X-6, we estimate an accretion rate of $\dot{m}_0 \sim 4$–9, depending on the assumed compact object mass in the range $5$–$15,M\odot$. This estimate places the source in a moderately super-Eddington regime, where the spherization radius (R$_{sph}$ = 270–850 km) lies significantly beyond the inner disk radius, R$_{in}$ = 30–130 km. This structure between $R_{\rm sph}$ and $R_{\rm in}$ is consistent with the expectation that radiation pressure drives strong outflows at larger radii, while the inner accretion disk remains relatively compact. The derived beaming factors ($b\sim 0.9$–2.3) further suggest that the observed L$_{X}$ of X-6 does not require strong geometric collimation, supporting a scenario in which the source is powered by nearly isotropic emission from a stellar-mass BH in a super-Eddington state \citep{2023ApJ...949...78G}. For the transient ULX X-7, the $L_{\mathrm{X}}$ versus $T_{\mathrm{in}}$ relation yields a slope of $\alpha = 2.54 \pm 0.55$, which is lower than the $\alpha = 4$ ($L \propto T^4$) scaling expected for a standard thin disk. However, according to the $\alpha$ range (0< $\alpha$ <4) defined by \cite{2025MNRAS.539.2064M}, X-7 lies within the regime corresponding to slim-disk behavior \citep{2000PASJ...52..133W,2004ApJ...601..428K}. Similarly, \citet{2016ApJ...816...60B} report that broadband spectral analyses of ULXs support the presence of optically thick, geometrically slim accretion disks, consistent with the observed parameters of X-7.

\subsubsection{Optical counterparts of ULXs X-6 and X-7}

The optical counterparts of ULXs X-6 and X-7 appear faint in both filters, consistent with the typical magnitudes observed for ULX optical counterparts. Because multiple possible counterparts are detected for both sources, we cannot place reliable constraints on their optical emission. Nevertheless, the CMDs allow us to derive some constraints on their ages and masses. However, all three candidate counterparts of X-6 occupy positions on the CMD that are clearly distinct from the field stars in the surrounding region (Figure \ref{F:cmds}), suggesting that they may belong to a different stellar population or have followed a different evolutionary path. The source X6a, which is relatively brighter than the other counterparts, is distinct from the isochrones on the CMD. The exceptionally blue color of X6a may be affected by observational limitations. Although we applied identical photometric procedures to all sources, the \textit{HST} data are limited and nonsimultaneous, leaving the optical variability of X6a unknown. If X6a is the true counterpart, its optical emission may be dominated by the accretion disk rather than by the donor star, which would produce a displaced position in the CMD. Deeper and simultaneous optical and X-ray observations are required to establish the true properties of X6a.

Based on the CMDs, X-6b, X-6c, X-7a, and X-7b have estimated ages of 30–40 Myr and masses in the range 7–9 M$_\odot$, consistent with an HMXB classification. Because continuous optical monitoring and simultaneous optical and X-ray observations are unavailable, the variability of these candidate counterparts remains uncertain. 
Consequently, we could not derive the F$_{X}$/F$_{optic}$ ratio for X-6 and X-7, which serves to constrain the nature of the sources through their X-ray-to-optical flux comparison \citep{1988ApJ...326..680M, 1991ApJS...76..813S}.

\subsection{NGC 1097 ULX-3}

In this study, we analyzed NGC 1097 ULX-3 in detail for the first time and identify it as an ULX based on its X-ray spectral and temporal properties, with additional observations in the optical and infrared bands. We detected ULX-3 in the five most recent  \textit{Chandra} observations between 21 May and 31 May 2025, with a stable L$_{X}$ $\sim$ (6--7) $\times$ 10$^{39}$ erg s$^{-1}$ and no significant variability on short timescales. The \textit{XMM-Newton} observation performed in 2023 shows a comparable luminosity level of $L_{\mathrm{X}} \sim 10^{40}$ erg s$^{-1}$. Long-term monitoring with \textit{Swift}/XRT between 2 August 2007 and 6 June  2025 reveals significant variability. During this 18-year period, L$_{X}$ of ULX-3 varied between $3 \times 10^{39}$ and $8 \times 10^{40}$ erg s$^{-1}$. Such long-term flux variation suggests that ULX-3 is a highly variable source, likely driven by significant changes in the accretion rate, accretion state transitions, or geometric effects such as variable absorption or accretion-disk precession. Overall, these results suggest that ULX-3 remains relatively stable on short timescales, as indicated by the recent \textit{Chandra} and \textit{XMM-Newton} observations, but exhibits obvious variability over long timescales spanning years to decades, as revealed by the \textit{Swift}/XRT monitoring.

 The X-ray spectrum of ULX-3 is best described by a two-component model consisting of a soft thermal blackbody and a hard power-law component (power-law plus bbody). The soft thermal component, with a temperature of \( kT \approx 0.15 \) keV, is consistent with the soft excess observed in many ULXs \citep{2009MNRAS.397.1836G,2013MNRAS.434.1702S} and is typically attributed to emission from an optically thick wind or outflow launched from a supercritical accretion disk \citep{2007MNRAS.377.1187P,2015MNRAS.454.3134M}. This component may correspond to the wind photosphere or to the outer, geometrically thick regions of the accretion flow. Such hard spectral slopes are commonly associated with the so-called HUL state in ULXs \citep{2009MNRAS.397.1836G,2013MNRAS.434.1702S,2014MNRAS.442.1054H}, in which the inner regions of the accretion flow may be partially visible through the wind funnel, allowing Comptonized emission from a hot corona or the innermost disk to dominate at higher energies. The combination of a low-temperature thermal component and a hard power-law tail supports a super-Eddington accretion scenario for ULX-3, likely involving a stellar-mass compact object surrounded by a geometrically and optically thick accretion structure \citep{2021MNRAS.505.5058P}.

To investigate spectral state transitions, we constructed hardness evolution and hardness intensity diagrams (Figure \ref{F:lc1097}, panels b and c). At high count rates, the source spectrum becomes softer, consistent with a thermal disk-dominated (high-soft) state, whereas at low count rates, the spectrum hardens, indicative of a corona- or Comptonization-dominated (low-hard) state. Such a correlation is a signature of accretion-powered variability, linking luminosity changes to transitions between spectral states. In Galactic XRBs, such transitions are well characterized and often correspond to changes between low-hard and high-soft states driven by variations in the mass accretion rate \citep{2006ARA&A..44...49R,2007A&ARv..15....1D}. In ULXs, however, spectral state transitions are more complex due to their potential super-Eddington luminosities and geometric beaming effects \citep{2009MNRAS.393L..41K}. The observed transitions in this source may reflect different accretion modes onto a compact object, which could be either a stellar-mass BH or an NS \citep{2007MNRAS.377.1187P, 2014Natur.514..202B,2015MNRAS.454.3134M,2018ApJ...856..128W}. Optically thick, radiatively driven outflows could obscure the inner hot regions of the disk, leading to softer spectra when observed at moderate inclination angles. In the case of harder spectra at even higher flux levels, changes in wind geometry, collimation, or the reemergence of the innermost accretion flow could indicate a transition in the accretion regime \citep{2012ApJ...752...18K}. Overall, the spectral state transitions observed in NGC 1097 ULX-3 reinforce the emerging picture that ULXs form a heterogeneous population, displaying diverse accretion behaviors that span the regime between Galactic XRBs and super-Eddington accretion, implying that ULX-3 could be powered either by a stellar-mass BH or an NS.

We identified a unique optical and NIR counterpart within the astrometric error radius of ULX-3. The optical counterpart is fainter than typical ULX counterparts, with F555W $\simeq m_V = 27.40$ mag ($M_V \simeq -4$ mag). In the F814W filter, the variability of the source is quantified as the difference between the faintest and brightest magnitudes measured over three epochs (2004–2019), yielding $\Delta m_{\rm F814W} \simeq 0.8$ mag and suggesting intrinsic variability in the optical emission of ULX-3. The variable X-ray emission is likely linked to the observed optical variability, suggesting that accretion processes dominate the optical emission.

In contrast, the NIR observations obtained between 27 December 2023 and 29 September 2024 reveal that the source remains nearly constant in magnitude. Furthermore, we can exclude a variable jet contribution to the NIR emission, since the infrared emission remains constant. Using simultaneous \textit{JWST}/NIRCam filters, we constructed the spectral energy distribution (SED) of the ULX counterpart and fit it with a blackbody model with a temperature of $\sim$ 3300 K (see Figure \ref{F:sedulx3}). The effective temperature derived is too high to be attributed to circumbinary material or dust \citep{2016ApJ...831...88D,2017ApJ...838L..17L,2019ApJ...878...71L,2023MNRAS.526.5765A} and instead falls within the expected range for RSGs. Adopting a blackbody temperature of 3300\,K, we used the relation $F_{\lambda} = \pi B_{\lambda}(T)\,(R/D)^2$, together with NIRCam photometry at a distance of 16.8 Mpc, to derive an RSG radius of $\sim200\,R_{\odot}$. The absolute magnitude of the counterpart is \(M_{277} = -8.9\) mag. In the NIR CMD, this value lies on the evolutionary tracks of \(\sim15{-}20\,M_{\odot}\) stars with ages of 7--12 Myr, indicating that its CMD position is consistent with the RSG evolutionary branch.

\section{Conclusions}\label{sec:5}

In this work, we report the identification and characterization of three new ULXs in NGC 4631 and NGC 1097. Our main findings can be summarized as follows.

NGC 4631 X-6 exhibits L$_{X}$ variability of about an order of magnitude and recurrent spectral variability. Spectral analyses support its classification as a stellar-mass BH candidate accreting in a super-Eddington regime. The $L_{\rm X}$–$T_{\rm in}$ relation follows the $L \propto T^4$ trend expected for a standard thin disk, supporting the interpretation of X-6 as a stellar-mass BH accretor. 

NGC 4631 X-7, a transient ULX, displays long-term L$_{X}$ variability and disk-dominated spectra with inner disk temperatures of 0.5–1 keV. Its photometric properties are consistent with a donor star of 7–9 M$\odot$ and an age of 30–40 Myr, supporting an HMXB scenario. Moreover, the $L \propto T^4$ trend expected for standard thin disks provides evidence for a slim disk accretion state. Finally, our analysis indicates that both X-6 and X-7 have properties consistent with normal HMXBs hosting BHs with masses of $\sim$ 20 M$\odot$.

NGC 1097 ULX-3 exhibits strong long-term variability, with L$_{X}$ changes up to a factor of $\sim$ 30. Spectral state transitions indicate that ULX-3 is a stellar-mass BH or NS accretor in a super-Eddington state. We identify a unique optical and NIR counterpart to ULX-3. Although the optical emission is highly variable, the NIR counterpart remains nearly constant. For the optical counterpart of ULX-3, the NIR SED is consistent with a blackbody temperature of $\sim3300$\,K and a stellar radius of $\sim$ 200 R$_{\odot}$, providing strong evidence that the source is an RSG donor star.\\
For the three ULXs in the two host galaxies, additional multiwavelength observations will be crucial to constrain their environments, donor stars, and accretion processes, ultimately leading to a clearer understanding of their nature.

\begin{acknowledgements}
This work is partially supported by the Bundesministerium f\"ur Wirtschaft und Energie through the Deutsches Zentrum f\"ur Luft- und Raumfahrt e.V. (DLR) under the grant 50 OR 2517.
LD acknowledges funding from the Deutsche Forschungsgemeinschaft (DFG, German Research Foundation) - Projektnummer 549824807. AA and YA acknowledge support provided by the TÜBİTAK through project number 124F004. We thank the referee for her/his careful reading of the manuscript and for the valuable comments, which helped us to improve the paper.
\end{acknowledgements}

\bibpunct{(}{)}{;}{a}{}{,}
\bibliographystyle{aa}
\bibliography{aa}

\begin{appendix}

\section{X-ray, optical, and NIR observations of the ULXs}

\begin{table}[ht!]
\caption{Log of NGC 4631 and NGC 1097 observations.}
\label{T:obs}
\begin{tabular}{cccccl}
\hline
Label|Filter & ObsID & Instrument & Exp.& Date \\
& & & (ks) & \\
\hline
&&NGC 4631 \\
\hline
 C1 & 797 & ACIS-S & 59.21 & 2000-04-16 \\
XM1 & 0110900201 & EPIC & 54.8 & 2002-06-28\\
XM2 & 0890710101 & EPIC & 33.0 & 2021-12-28\\
C2 & 25777 & ACIS-S & 29.03 & 2022-01-22 \\
C3 & 25220 & ACIS-S & 22.77 & 2022-08-02 \\
C4 & 26484 & ACIS-S & 18.82 & 2022-08-02 \\
C5 & 26485 & ACIS-S & 20.80 & 2022-08-05 \\
C6 & 26486 & ACIS-S & 14.88 & 2022-08-06 \\
C7 & 26487 & ACIS-S & 14.88 & 2022-08-07 \\
C8 & 25782 & ACIS-S & 30.66 & 2023-01-29 \\
C9 & 25780 & ACIS-S & 11.92 & 2023-06-16 \\
C10 & 25779 & ACIS-S & 19.82 & 2023-07-04 \\
C11 & 25778 & ACIS-S & 19.81 & 2023-07-04\\
C12 & 25781 & ACIS-S & 13.60 & 2023-07-18\\
XM3 & 0943790101 & EPIC & 104.20 & 2025-07-08\\
\hline
F606W & j8r331010 & ACS/WFC & 0.68 & 2003-08-03\\
F814W & j8r331020 & ACS/WFC & 0.70 & 2003-08-03\\
F606W & jc9l04010 & ACS/WFC & 2.40 & 2014-01-21\\
F814W & jc9l04020 & ACS/WFC & 2.53 & 2014-01-21\\
\hline
&&NGC 1097 \\
\hline
CH1 & 2339 & ACIS-S & 5.34 & 2001-01-28 \\
CH2 & 1611 & ACIS-S & 5.34 & 2001-01-28 \\
XMM1 & 922490201 & EPIC & 33 & 2023-12-22\\
CH3 & 30142 & ACIS-S & 9.94 & 2025-05-21 \\
CH4 & 30937 & ACIS-S & 9.94 & 2025-05-21 \\
CH5 & 30938 & ACIS-S & 9.94 & 2025-05-21 \\
CH6 & 29637 & ACIS-S & 14.71 & 2025-05-31 \\
CH7 & 30948 & ACIS-S & 14.71 & 2025-05-31 \\
\hline
F814W & j8mx03fsq & ACS/WFC & 0.12 & 2004-06-05 \\
F336W & icc501021 & WFC3/UVIS & 1.43 & 2014-02-20 \\
F438W & icc501041 & WFC3/UVIS & 0.81 & 2014-02-20 \\
F814W & icc501050 & WFC3/UVIS & 0.70 & 2014-02-20 \\
F814W & jdxk38010 & ACS/WFC & 2.15 & 2019-05-28 \\
F275W & idxr58040 & WFC3/UVIS & 2.3 & 2020-08-14 \\
F555W & idxr58030 & WFC3/UVIS & 2.30 & 2020-08-14 \\

F115W & jw03707 & NIRCam/IMAGE & 0.22 & 2023-12-27 \\
F115W & jw04793 & NIRCam/IMAGE & 0.86 & 2024-09-29 \\
F200W & jw04793 & NIRCam/IMAGE & 0.86 & 2024-09-29 \\
F300M & jw03707 & NIRCam/IMAGE & 0.22 & 2023-12-27 \\
F335M & jw03707 & NIRCam/IMAGE & 0.86 & 2023-12-27 \\
F335M & jw04793 & NIRCam/IMAGE & 1.63 & 2024-09-29 \\
F360M & jw04793 & NIRCam/IMAGE & 1.63 & 2024-09-29 \\
F444W & jw04793 & NIRCam/IMAGE & 1.63 & 2024-09-29 \\
F277W & jw04793 & NIRCam/IMAGE & 0.86 & 2024-09-29 \\
\hline

\end{tabular}
\\ Notes: NGC 4631, with Target IDs 00082263 and 00084441, was observed 24 times between 2013-11-08 and 2021-02-13. For NGC 1097 ULX-3, we analyzed 84 \textit{Swift}/XRT PC observations (Target ID 00045597, 2011 Aug 05–2025 Jun 06) and 2 additional observations (Target ID 00036582, 2007 Aug 02–Dec 16).
 \\
\end{table}

\end{appendix}

\end{document}